\documentclass[12pt]{iopart}

\usepackage{iopams}  
\usepackage{subcaption}
\usepackage{widetext,graphicx,amsfonts,xcolor}
\usepackage{hyperref}
\expandafter\let\csname equation*\endcsname\relax
\expandafter\let\csname endequation*\endcsname\relax
\usepackage{amsmath,braket}
\begin{document}

\title{QKD in the NISQ era: enhancing secure key rates via quantum error correction}

\author{Shashank Kumar Ranu$^{1,2}$, Anil Prabhakar$^1$, 
Prabha Mandayam$^2$}

\address{$^1$Department of Electrical Engineering, Indian Institute of Technology Madras, Chennai, India}
\address{$^2$Department of Physics, Indian Institute of Technology Madras, Chennai, India}
\eads{\mailto{ee16s300@ee.iitm.ac.in}, \mailto{anilpr@ee.iitm.ac.in, \mailto{prabhamd@physics.iitm.ac.in}}}

\vspace{10pt}
\begin{indented}
\item[]\today
\end{indented}

\begin{abstract}
 Error mitigation is one of the key challenges in realising the full potential of quantum cryptographic protocols. Consequently, there is a lot of interest in adapting techniques from quantum error correction (QEC) to improve the robustness of quantum cryptographic protocols. In this work, we benchmark the performance of different QKD protocols on noisy quantum devices, with and without error correction. We obtain the secure key rates of BB84, B92 and BBM92 QKD protocols over a quantum channel that is subject to amplitude-damping noise. We demonstrate, theoretically and via implementations on the IBM quantum processors, that B92 is the optimal protocol under amplitude-damping and generalized amplitude-damping noise. We then show that the security of the noisy BBM92 protocol crucially depends on the type and the mode of distribution of an entangled pair. Finally, we implement an error-corrected BB84 protocol using dual-rail encoding on a noisy quantum processor, and show that the dual-rail BB84 implementation outperforms the conventional BB84 in the presence of noise. Our secure key rate calculation also takes into account the effects of \textsc{cnot} imperfections on the error rates of the protocols.
\end{abstract}

\section{Introduction}\label{sec: Intro}
Quantum key distribution (QKD) offers the promise of secure communication between two parties, Alice and Bob, in the presence of an eavesdropper, Eve. When Eve attempts to steal information from the quantum channel, she also inevitably introduces disturbances in the channel and reveals herself. Since the first proposal by Bennett and Brassard in 1984 \cite{bennett2020quantum}, there have been many advances, both theoretically and experimentally \cite{gisin2002quantum,pirandola2019advances,Xu_2020}. So far, QKD protocols have been implemented over thousands of kilometers over free-space channels \cite{liao2017satellite}, hundreds of kilometers over fiber-optical channels \cite{pirandola2019advances,Xu_2020}, and tens of meters over underwater quantum communication channels \cite{feng2021experimental}. QKD key rates are affected by scattering, absorption, damping, and other noise models prevalent in these quantum channels. Amplitude-Damping (AD) and Generalized Amplitude-Damping (GAD) are two such noise models that degrade the performance of a QKD system, especially in turbulent media such as air and seawater.  
	
In parallel, the area of quantum computing has also progressed dramatically, culminating in the demonstration of quantum supremacy by various groups \cite{arute2019quantum,Zhong1460,wu2021strong}. Established technology giants (IBM, Microsoft, Google, Honeywell, etc.) and startups (Rigetti, IonQ, Xanadu, etc.) are in the race to build a fully fault-tolerant and scalable quantum computer. However, the current generation of quantum computers are Noisy Intermediate Scale Quantum (NISQ) devices with noisy gates, qubit readout errors, and small coherence times \cite{preskill2018quantum}. Benchmarking these NISQ devices forms a critical step towards coming up with the practical applications of these quantum processors.
	
Effects of AD noise on the BB84 QKD protocol have been studied in the asymptotic as well as finite-key regime \cite{Watanabe_2008,sano2010secure}. However, the performance of the BB84 protocol in the presence of GAD noise has not been studied. Entanglement-based protocols such as the BBM92 protocol form a vital subclass of QKD protocols. Effects of AD noise on the Bell states have been extensively studied in the context of teleportation fidelity between two parties \cite{Som_teleportation2002}, but its effects on the bit and the phase error rates of an entanglement-based QKD protocol are yet to be quantified.  

In this work, we merge ideas from two different areas of quantum technologies, namely, QKD and quantum computing. 
We employ a quantum processor to mimic the amplitude-damping channel and implement QKD protocols on it. Such an implementation on a quantum processor helps us observe the effects of AD noise on the performance of QKD protocols without having to physically implement the protocol over a noisy, long-distance quantum channel. Our results will also help us design efficient QKD protocols over turbulent channels such as seawater and free-space. The insights we gain from this study will also help in the eventual design and characterization of quantum memories to be used as quantum repeaters, in the presence of AD noise.

We invoke techniques from quantum error correction to mitigate the effects of amplitude-damping noise on the secure key rates of QKD protocols. Recently, there have been preliminary works studying the performance of encoded quantum repeater-based QKD protocols, where the  well-known three-qubit repetition code has been employed for encoding quantum information \cite{Jing_2020,jing2022quantum}. Here, in a deviation from the standard approach of using stabilizer codes to correct for \emph{arbitrary} noise, we rely on  noise-adapted quantum error correction~\cite{jayashankar2022} to improve the secure key rates of QKD protocols over an amplitude-damping channel.       

One of the simplest error-detecting codes tailored to protect against amplitude-damping noise is the dual-rail code, involving encoding a logical qubit in just a pair of physical qubits~\cite{nielsen2002quantum}. In previous work, dual-rail encoding has been used to correct readout asymmetry in a BB84 implementation on the IBM quantum processors \cite{Zhukov_2018}, where the \textsc{swap} gate was used to realise a quantum channel between Alice and Bob. 
The \textsc{swap} gates are implemented using \textsc{cnot} gates, thereby making depolarizing error the dominant noise in this BB84 implementation of \cite{Zhukov_2018}. Moreover, the dual-rail encoder and decoder used in~\cite{Zhukov_2018} consist of $4$ \textsc{cnot} gates, leading to a high intrinsic error rate. Furthermore, the results in~\cite{Zhukov_2018} do not address the effects of state preparation errors, gate imperfections, channel noise, and qubit readout errors of the NISQ device on the phase and bit error rates of their dual-rail-based protocol. 

In our work, we present two variants of the dual-rail-encoded BB84 protocol and introduce a theoretical framework to describe the effects of \textsc{cnot} imperfections as well as damping noise on the performance of our dual-rail BB84 implementations. First, we use an error detection scheme presented in \cite{bose2005} to reduce the number of \textsc{cnot}s in the dual-rail BB84 implementation. We compare the performance of this error-detected scheme against the conventional BB84 protocol under AD noise. Our study also helps to identify the optimal prepare-and-measure QKD protocol under AD noise. 

Furthermore, we also address some practical questions in the context of the BBM92 protocol using our theoretical framework and implementations. In particular, we identify the Bell-EPR states that are most suited for the BBM92 protocol in the presence of AD noise. We also determine an optimal way to share an entangled pair between Alice and Bob for the BBM92 protocol. Finally, we estimate the effects of channel asymmetry, that is, different damping probabilities of Alice and Bob's channel, on the performance of the BBM92 protocol. 

The rest of this paper is organized as follows. Sec.~\ref{sec:prelims} of this work gives an overview of AD and GAD noise. We also discuss the robustness of the dual-rail encoding against both noise models. Sec.~\ref{sec: AD on key rate analytical} describes the effects of AD and GAD noise on the secure key rates of different QKD protocols. We also consider various practical imperfections that affect QKD protocols under AD noise. We next study the performance of dual-rail encoded BB84 against AD noise in Sec.~\ref{sec:dual-rail}. Assuming non-ideal quantum circuits, we also estimate the effects of gate imperfections on the performance of the dual-rail BB84 protocol. Sec.~\ref{sec:Implementation} of our paper describes the implementation of different QKD protocols on IBM quantum processors. Finally, we summarize and discuss future outlook in Sec.~\ref{sec:Conclusion}.

\section{Preliminaries}\label{sec:prelims}
\subsection{Amplitude-damping channel}\label{subsec: AD}
The amplitude-damping (AD) channel models the decay of an excited state of a two-level atom due to the spontaneous emission of a photon. 
Concretely, the interaction between a two-level atom and the electromagnetic field is described using the Jaynes-Cummings Hamiltonian ($H_{\text{JC}}$). Time evolution of the joint system is described by the unitary operator $U=e^{-iH_{\text{JC}}t}$. Fixing the detuning parameter of the Jaynes-Cummings interaction as $0$ and tracing over the field leads to a decay probability for the two-level atom that has an exponential dependence on time \cite{nielsen2002quantum}.

Let $\ket{0}_A$ denote the ground state of a two-level atom and $\ket{1}_A$ represent the excited state of the atom. We assume the “environment” described by the modes of the electromagnetic field to be in the vacuum state $\ket{0}_E$. Let $\gamma(t)$ denote the decay probability, with $0 \leq \gamma \leq 1$.  After a time $t$, the excited state decays to the ground state with a probability $\gamma (t)$, and the atom emits a photon. Hence, the environment makes a transition from the state $\ket{0}_E$ (“no photon”) to the state $\ket{1}_E$
(“one photon”). This evolution is described as \cite{nielsen2002quantum},
\begin{align}
    \ket{0}_A\otimes \ket{0}_E &\to \ket{0}_A\otimes \ket{0}_E, \nonumber \\  
    \ket{1}_A\otimes\ket{0}_E &\to \sqrt{1-\gamma (t)}\ket{1}_A\otimes \ket{0}_E +\sqrt{\gamma (t)}\ket{0}_A\otimes \ket{1}_E . \label{eq:a}  
\end{align}
This disspiative behavior is captured by a single-qubit operator of the form,
\begin{equation}
    A_1^{\text{AD}}=\sqrt{\gamma(t)}\ket{0}\bra{1}.
\end{equation} \label{eq:b}
Note that we will drop the dependence of $\gamma$ on time henceforward for notational simplicity. $A_1^{\text{AD}}$ annihilates the ground state and causes the excited state to decay to the ground state. However, the Kraus operator $A_1^{\text{AD}}$ alone does not specify a physical map, since $A_1^{\text{AD}\dagger}A_1^{\text{AD}}=\gamma\ket{1}\bra{1}$. The Kraus operators of any channel should satisfy the condition $\sum_i A_i^\dagger A_i=I$. We satisfy this completeness condition by choosing another operator $A_0^{\text{AD}}$ such that,
\begin{equation}
    A_0^{\text{AD}\dagger}A_0^{\text{AD}}= I-A_1^{\text{AD}\dagger}A_1^{\text{AD}} = \ket{0}\bra{0}+\sqrt{1-\gamma}\ket{1}\bra{1}.
\end{equation}
Thus, the operators $A_0^{\text{AD}}$ and $A_1^{\text{AD}}$ are valid Kraus operators for the AD channel, and are represented in matrix form as,
\begin{equation}
A_0^{\text{AD}} = \begin{pmatrix}
1 & 0\\
0 & \sqrt{1-\gamma}
\end{pmatrix}, \quad A_1^{\text{AD}} = \begin{pmatrix}
0 & \sqrt{\gamma}\\
0 & 0
\end{pmatrix}.\label{eq:kraus}
\end{equation}

\subsection{Generalized amplitude-damping channel}
The GAD channel models the dissipation effects due to the interaction of a system with a purely thermal bath at finite temperature. GAD noise is one of the primary sources of noise in superconducting quantum processors and in linear optical systems with low-temperature background noise \cite{PhysRevA.95.042342,chirolli2008decoherence}. The Kraus representation of the GAD channel is, 
\begin{equation}
    \Lambda (\rho)=\sum_{i=0}^{i=3}A_i^{\text{GAD}}\rho A_i^{\dagger\text{GAD}},
\end{equation}\label{eq:GAD}
where the $2 \times 2$ matrix representation of the Kraus operators of GAD channel are
\begin{eqnarray}
&& A_0^{\text{GAD}}=\sqrt{p}\begin{pmatrix}
1 & 0\\
0 & \sqrt{1-\gamma}
\end{pmatrix}, \; 
A_1^{\text{GAD}}=\sqrt{p}\begin{pmatrix}
0 & \sqrt{\gamma}\\
0 & 0
\end{pmatrix},\nonumber \\
&& A_2^{\text{GAD}}=\sqrt{1-p} \begin{pmatrix}
\sqrt{1-\gamma} & 0\\
0 & 1
\end{pmatrix}, \nonumber \\
&& A_3^{\text{GAD}}=\sqrt{1-p}\begin{pmatrix}
0 & 0\\
\sqrt{\gamma} & 0
\end{pmatrix}.
\label{eq:GAD Kraus}
\end{eqnarray}
Here, $\gamma$ is the damping parameter and $0\leq p \leq 1$. The action of Kraus operators of the GAD channel on the computational basis vectors $\{\ket{0},\ket{1}\}$ can be expressed as,
\begin{eqnarray}
&& A_0^{\text{GAD}}\ket{0}=\sqrt{p}\ket{0}, \; A_0^{\text{GAD}}\ket{0}=\sqrt{p}\sqrt{1-\gamma}\ket{1}, \nonumber \\  
&& A_1^{\text{GAD}}\ket{0}=0, \; A_1^{\text{GAD}}\ket{1}=\sqrt{p}\sqrt{\gamma}\ket{0}, \nonumber \\
&& A_2^{\text{GAD}}\ket{0}=\sqrt{1-p}\sqrt{1-\gamma}\ket{0}, \; A_2^{\text{GAD}}\ket{1}=\sqrt{1-p}\ket{1}, \nonumber \\
&& A_3^{\text{GAD}}\ket{0}=\sqrt{1-p}\sqrt{\gamma}\ket{1}, \, A_3^{\text{GAD}}\ket{1}=0.
\end{eqnarray}\label{eq:GAD Kraus computation}
Note that we can obtain the Kraus operators of the amplitude-damping channel from Eq.~\eqref{eq:GAD Kraus} for $p=1$. As the name suggests, the GAD channel generalizes the AD channel for a bath at a finite temperature. Hence, GAD noise leads to transitions from $\ket{0} \to \ket{1}$ as well as $\ket{1}\to \ket{0}$. Such a transformation may happen in practical QKD systems due to polarization fluctuations.

\subsection{Robustness of dual-rail encoding against amplitude-damping noise}\label{subsec:dual-rail}
\begin{figure}[h]
\centerline{\includegraphics[width=0.35\textwidth,height=\textheight,keepaspectratio]{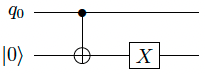}}

     \caption{Circuit for dual-rail qubit. Here, $q_0$ can be in any arbitrary state.}
\label{fig:encoder}
\end{figure}

We now briefly review the dual-rail encoding and how it acts as an error-detecting code against single-qubit amplitude-damping noise~\cite{nielsen2002quantum}. Dual-rail encoding maps a qubit onto a two-qubit subspace as shown below.
\begin{equation}
\ket{0} \to \ket{01} \equiv \ket{0}_L, \quad \ket{1} \to \ket{10} \equiv \ket{1}_L. \label{eq:dual}
\end{equation}
The circuit shown in Fig.~\ref{fig:encoder} implements such an encoding.
Using Eq.~\eqref{eq:kraus} we can write the Kraus operators for the dual-rail qubits as,
\begin{eqnarray}
M_0&=& A_0^{\text{AD}}\otimes A_0^{\text{AD}} \nonumber \\
M_1&=& A_0^{\text{AD}}\otimes A_1^{\text{AD}} \nonumber \\
M_2&=& A_1^{\text{AD}}\otimes A_0^{\text{AD}} \nonumber \\
M_3&=& A_1^{\text{AD}}\otimes A_1^{\text{AD}} 
\end{eqnarray}
And the action of these Kraus operators on the logical qubits can be described as shown below.
\begin{equation}
	\begin{split}
&M_0\ket{01}=\sqrt{1-\gamma}\ket{01},\\
&M_1\ket{01}=0,\\
&M_2\ket{01}=\sqrt{\gamma}\ket{00},\\
&M_3\ket{01}=0,
\end{split} 
\quad
\begin{split}
&M_0\ket{10}=\sqrt{1-\gamma}\ket{10},\\
&M_1\ket{10}=\sqrt{\gamma}\ket{00},\\
&M_2\ket{10}=0,\\
&M_3\ket{10}=0.
\end{split}
\label{eq:dualkraus}
\end{equation}
It can be seen from Eq.~\eqref{eq:dualkraus} that the Kraus operators of the AD channel either annihilate the logical qubits or map it to an orthogonal subspace spanned by $\ket{00}$. This mapping serves as the basis for the robustness of the dual-rail qubit against AD noise. 

\begin{figure}[h]
\centerline{\includegraphics[width=0.7\textwidth,height=\textheight,keepaspectratio]{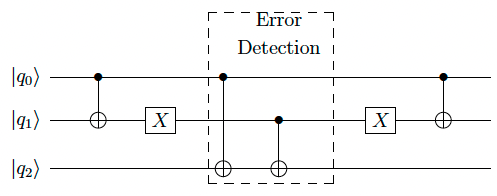}}
     \caption{Circuit for detecting amplitude-damping error.}
				\label{fig:detection}
			\end{figure}

We use the circuit shown in Fig.~\ref{fig:detection} to detect AD errors. From Eq.~\eqref{eq:dual}, we see that the dual-rail qubits have an odd parity, which can be estimated by using a pair of \textsc{cnot}s. Hence, a measurement outcome of $\ket{0}$ for the ancilla qubit indicates a change in the parity of the dual-rail qubits, thereby detecting an AD error. In \ref{appendix:ancilla-based PS}, we show via explicit calculation that the error-detection circuit of Fig.~\ref{fig:detection} indeed detects the occurrence of a single AD error on any arbitrary qubit. We further show in Secs.~\ref{sec:dual-rail} and ~\ref{sec:GAD_dual} that the error-detection circuit of Fig.~\ref{fig:detection} can also detect a fraction of GAD errors.

\subsection{Prepare-and-measure QKD protocols}\label{subsec:bb84}
In this subsection, we give a brief overview of the two well-known prepare-and-measure QKD protocols - BB84 and B92. In the BB84 protocol, a sender (Alice) transmits single photons polarized along horizontal/vertical and diagonal/anti-diagonal axes. The receiver (Bob) measures these encoded photons randomly in either $\sigma_z$ basis or $\sigma_x$. The raw key generation is followed by classical post-processing steps to obtain a secure key.
The secure key rate is given by \cite{shor2000simple}
\begin{equation}
    l_{\text{sec}}^{\text{BB84}}=l_{\text{sift}}[1-h(e_b)-h(e_p)]. \label{eq:key}
\end{equation}
Here, $l_{\text{sec}}$ is the secure key length, $l_{\text{sift}}$ is the sifted key length, $e_b$ represents the quantum bit error rate QBER) and $e(p)$ is the phase error rate of the BB84 protocol. Due to the symmetry of the BB84 protocol, its phase error rate equals its bit error rate (QBER). Hence, secure key generation becomes impossible once QBER exceeds $11\%$ \cite{shor2000simple}.

In the B92 protocol, Alice encodes key information using a pair of non-orthogonal states \cite{B921}. Like BB84, B92 also uses two basis, but just one eigenstate from each of the bases. Alice encodes classical bit $0$ in the rectilinear basis and $1$ as $\ket{+}$ in the diagonal basis. The measurement setup of Bob is same as in BB84. He randomly measures the encoded key in linear and diagonal basis. Measurement is followed by the usual steps of sifting, error estimation, error correction and privacy amplification. Here, Alice and Bob establish a secure key using fewer encodings, thereby leading to a simpler experimental setup compared to the BB84 protocol.

\subsection{BBM92 QKD}\label{subsec:bbm92}
Unlike BB84 and other prepare-and-measure QKD protocols, entanglement-based (EB) QKD protocols do not require an active basis choice during state preparation \cite{E91,BBM92}. THe BBM92 QKD protocol uses entanglement as a resource to generate a shared secret key between Alice and Bob. This requires Alice and Bob to first share a maximally entangled state of two qubits, often referred to as a \emph{Bell-EPR pair}. Alice and Bob measure their respective halves of the EPR pair randomly in the $X$ or $Z$ basis to generate a raw key. Similar to prepare-and-measure QKD protocols, Alice and Bob use classical communication to do sifting, error estimation, error correction, and privacy amplification. The security of BBM92 relies on monogamy of entanglement, which prohibits Eve from establishing correlations with Alice and Bob's qubits without introducing errors in their measurement statistics.

\section{Secure key rates for QKD over noisy quantum channels}\label{sec: AD on key rate analytical}
In this section, we quantify the effects of AD and GAD noise channels on the secure key rate of three well known QKD protocols, namely, BB84, B92 and BBM92. To make our analysis more comprehensive, we also take into account the effect of measurement errors. 

\subsection{Effect of amplitude-damping noise on the secure key rate of BB84}\label{subsec: AD on BB84 key rate analytical}
We first study the effects of AD noise on the performance of the BB84 protocol. Bob's projective measurement in the $\{\ket{0},\ket{1}\}$ basis is defined as,
\begin{equation}
P_0=\ket{0}\bra{0}, \quad P_1=\ket{1}\bra{1}.
\label{eq:POVM_Bob}
\end{equation}
Similarly, we use $P_{\pm}$ to denote Bob's measurement operators for the $X$ basis. We use Eq.~\eqref{eq:key} to calculate the secure key rate of the BB84 protocol, where the error rates $e_b$ and $e_p$ are now obtained by assuming that the states sent by Alice are subject to AD noise, as, 
\begin{eqnarray}
e_b^{\text{AD}}&=&\frac{1}{2}\big(\text{Tr}\big(P_1 \Lambda^{\text{AD}}(\ket{0}\bra{0})\big)+ \text{Tr} \big((P_0 \Lambda^{\text{AD}}(\ket{1}\bra{1})\big)\big).\label{eq:bit error rate bb84}  \end{eqnarray}
\begin{eqnarray}
e_p^{\text{AD}}&=&\frac{1}{2}\big(\text{Tr} \big(P_- \Lambda^{\text{AD}}(\ket{+}\bra{+})\big)+\text{Tr} \big(P_+ \Lambda^{\text{AD}}(\ket{-}\bra{-})\big)\big).\label{eq:phase error rate bb84}
\end{eqnarray}
Here, $\Lambda^{\text{AD}}(.)$ represents action of the AD channel on the BB84 states. Note that bit error arises when Alice sends $\ket{0}$ ($\ket{1}$), but Bob decodes the received qubit as $\ket{1}$ ($\ket{0}$). Such a mismatch between Alice's state and Bob's measurement may occur due to noise, eavesdropping or practical imperfections such as read-out errors. In our analysis, we assume that the bit error arises solely due to AD noise, as characterized by $\Lambda^{\text{AD}}(.)$ in Eq.~\eqref{eq:bit error rate bb84}. Similarly, phase error arises because Bob obtains $\ket{-}$ ($\ket{+}$) as his measurement outcome when Alice sends $\ket{+}$ ($\ket{-}$). Once again we restrict our attention to phase errors arising due to amplitude-damping noise and this rate is quantified in Eq.~\eqref{eq:phase error rate bb84}.  

We now give the final expressions for the bit and phase error rates obtained via Eqs.~\eqref{eq:bit error rate bb84} and~\eqref{eq:phase error rate bb84} and refer to \ref{appendix:AD on BB84 states} for the details.  
 \begin{equation}
    e_b^\text{{AD}}= \frac{\gamma}{2}, \quad e_p^{\text{AD}}=\frac{1}{2}\left(1+\sqrt{1-\gamma}\right).
 \end{equation}
 Hence, the secure key rate of the BB84 protocol under AD noise is given by,
 \begin{eqnarray}
 R_\text{BB84}^{\text{AD}}&=&1-h\left(\frac{\gamma}{2}\right)-h\left(\frac{1}{2}\left(1+\sqrt{1-\gamma}\right)\right).\label{eq:AD BB84 key rate}
 \end{eqnarray}
Finally, we have also quantified the effects of qubit read-out errors on Bob's projective measurement and thereby on the secure key rate of the BB84 protocol in \ref{appendix:AD on BB84 states}. 

\subsection{Effect of GAD noise on the secure key rate of BB84}\label{subsec:GAD on BB84}

We next quantify the effect of GAD channel on the performance of the BB84 protocol, and compare its performance with the BB84 protocol under AD noise.  Bob's projective measurement to states $\ket{0}$ and $\ket{1}$ are as shown in Eq.~\eqref{eq:POVM_Bob}. We use the Shor-Preskill security proof \cite{shor2000simple} to obtain the secure key rate of the BB84 protocol under GAD noise (see Eq.~\eqref{eq:key}). We calculate the error rates, $e_b$ and $e_p$, for the BB84 protocol under GAD noise as,
\begin{eqnarray}
e_b^{\text{GAD}}&=&\frac{1}{2}\big(\text{Tr}\big(P_1 \Lambda^{\text{GAD}}(\ket{0}\bra{0})\big)+ \text{Tr} \big((P_0 \Lambda^{\text{GAD}}(\ket{1}\bra{1})\big)\big),  \label{eq:GAD bit error rate bb84}
\end{eqnarray}
\begin{eqnarray}
e_p^{\text{GAD}}&=&\frac{1}{2}\big(\text{Tr} \big(P_- \Lambda^{\text{GAD}}(\ket{+}\bra{+})\big)+\text{Tr} \big(P_+ \Lambda^{\text{GAD}}(\ket{-}\bra{-})\big)\big),
\label{eq:GAD phase error rate bb84}
\end{eqnarray}
where $\Lambda^{\text{GAD}}(.)$ represents the action of the GAD channel on the BB84 states and is evaluated explicitly in \ref{appendix:GAD on BB84 states}.

 We then obtain the bit and phase error rates of the BB84 protocol under GAD noise as, 
 \begin{equation}
 e_b^\text{{GAD}}= \frac{\gamma}{2}, \quad e_p^{\text{GAD}}=\frac{1}{2}\left(1+\sqrt{1-\gamma}\right).
 \end{equation}
 Hence, key rate of the BB84 under GAD noise is
 \begin{eqnarray}
 R_\text{BB84}^{\text{GAD}}&=&1-h\left(\frac{\gamma}{2}\right)-h\left(\frac{1}{2}\left(1+\sqrt{1-\gamma}\right)\right).\label{eq: GAD bb84 key rate}
 \end{eqnarray}
 We observe from Eq.~\eqref{eq:AD BB84 key rate} and Eq.~\eqref{eq: GAD bb84 key rate} that both AD and GAD noise have an identical effect on the secure key rate of the BB84 protocol. 
 
  \subsection{Effects of amplitude-damping noise on the B92 protocol}
 We next present our theoretical results on the effects of AD noise on the B92 protocol. In this case, using a similar argument as in the case of BB84, the error rates are given by,
\begin{equation}
e_b = \text{Tr}\left(P_1 \Lambda^{\text{AD}}(\ket{0}\bra{0})\right), \quad e_p = \text{Tr} \left(P_{-} \Lambda^{\text{AD}}(\ket{+}\bra{+})\right).
\label{eq:error rate b92}
\end{equation}    
Here, $\Lambda^{\text{AD}}(.)$ represents the action of the AD channel on the B92 states and $P_{1}$ and $P_{-}$ denote projection operators for Bob's measurement as defined in Eq.~\eqref{eq:POVM_Bob}. As before, we refer to ~\ref{appendix:AD on BB84 states} for detailed calculations on action of AD channel on $\ket{0}$ and $\ket{+}$. We have also shown the effects of Bob's imperfect projective measurements on the security of B92 protocol in ~\ref{appendix:AD on BB84 states}.  For conciseness, we directly state the final expressions for the error rates of the B92 protocol under AD noise.  
 \begin{equation}
e_b= 0, \quad e_p=\frac{1}{2}\left(1+\sqrt{1-\gamma}\right).
 \end{equation}
 Hence, the secure key rate of B92 under AD noise is given by,
 \begin{eqnarray}
 R_\text{B92}&=&1-h\left(\frac{1}{2}\left(1+\sqrt{1-\gamma}\right)\right).
 \end{eqnarray}
 
 \subsection{Effects of amplitude damping on BBM92}
 
Recall that the BBM92 protocol requires Alice and Bob to share a maximally entangled state of two qubits. The original paper that introduced BBM92 uses the maximally correlated Bell-EPR state ($\rho^{\vert\vert}$) for key generation \cite{BBM92}, where,
\begin{equation}
  \rho^{\vert\vert}=  \frac{1}{\sqrt{2}}\left(\ket{00}_{AB}+\ket{11}_{AB}\right).
\end{equation}
However, many experimental realizations of BBM92 use a maximally anti-correlated Bell-EPR pair ($\rho^{\perp}$) \cite{erven2012experimental}, where,
\begin{equation}
    \rho^{\perp}=\frac{1}{\sqrt{2}}\left(\ket{01}_{AB}-\ket{10}_{AB}\right).
\end{equation}
 
In this subsection, we study the effects of AD noise on different pairs of maximally entangled states, and hence on the secure key rate of the BBM92 QKD protocol. We then quantify the security key rate of BBM92 protocol under AD noise. Finally, we investigate the effects of channel asymmetry on BBM92 security in this subsection.
 \subsubsection{Effect of amplitude-damping channel on different Bell-EPR pairs:}\label{subsub:EPR pairs}

It has been noted previously, that AD noise affects the Bell states differently, leading to different teleportation fidelities \cite{Som_teleportation2002}. We now study this in the context of QKD. Specifically, we are interested in the bit and phase error rates of BBM92 protocols using different Bell states in the presence of AD noise.

We use Eq.~\eqref{eq:key} to calculate the secure key rate of the BBM92 protocol using $\rho^{\vert\vert}$ as an entangled resource. The error rates $e_b$ and $e_p$ for such a protocol are obtained as,
\begin{eqnarray}
&& e_b=\text{Tr}\left(P_0^{\text{A}} P_1^{\text{B}} \Lambda^{\text{AD}}(\rho^{\parallel}_{AB})\right) + \text{Tr} \left(P_1^{\text{A}} P_0^{\text{B}} \Lambda^{\text{AD}}(\rho^{\parallel}_{AB})\right), \nonumber \\ 
&& e_p=\text{Tr} \left(P_+^{\text{A}} P_-^{\text{B}} \Lambda^{\text{AD}}(\rho^{\parallel}_{AB})\right) + \text{Tr} \left(P_-^{\text{A}} P_+^{\text{B}} \Lambda^{\text{AD}}(\rho^{\parallel}_{AB})\right),
\label{eq:error rate bbm92}
\end{eqnarray}
where $\Lambda()^{\text{AD}}$ represents the action of the AD channel on EPR pairs of type $\rho^{\vert\vert}$. We assume that a third party, Charlie, generates the EPR pairs, so the initial state is, 
\begin{equation}
   \rho_{\text{Charlie}}^{\parallel}=\frac{1}{2}\begin{pmatrix}
1 & 0 & 0 & 1\\
0 & 0 & 0 & 0 \\
0 & 0 & 0 & 0 \\
1 & 0 & 0 & 1
\end{pmatrix}. \label{eq:cor den mat}
\end{equation}
When Charlie sends one half of the EPR pair each to Alice and Bob, the AD channel acts on the state $\rho^{\vert\vert}$ as shown below.
\begin{eqnarray}
\rho_{AB}^{\vert \vert} &=&\sum_{i,j=0}^{i,j=1}\Big(A_i^{\text{AD}}\otimes A_j^{\text{AD}} \, \left(\rho_{\text{Charlie}}^{\vert\vert}\right) \, A_i^{\text{AD}^\dagger}\otimes A_j^{\text{AD}\dagger}\Big).
\end{eqnarray}
The resultant density matrix is,
\begin{equation}
   \rho_{AB}^{\parallel}=\frac{1}{2}\begin{pmatrix}
1+\gamma^2 & 0 & 0 & 1-\gamma\\
0 & \gamma(1-\gamma) & 0 & 0 \\
0 & 0 & \gamma(1-\gamma) & 0 \\
1-\gamma & 0 & 0 & (1-\gamma)^2
\end{pmatrix}. \label{eq:bbm92_corr}
\end{equation}
Using Eq.~\eqref{eq:error rate bbm92}, we find that the bit and phase error rates for the BBM92 protocol using $\rho^{\vert \vert}$ as the EPR resource state, are given by,
\begin{equation}
e_b=\gamma(1-\gamma), \quad e_p=\frac{\gamma}{2}.
\end{equation}

\begin{figure}[h!]
				\centerline{\includegraphics[width=0.5\textwidth,height=\textheight,keepaspectratio]{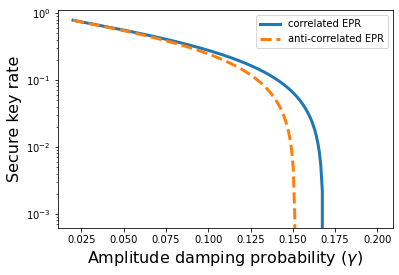}}
				\caption{Comparison of secure key rates BBM92 using correlated and anti-correlated EPR pairs.}
				\label{fig:theory_EPR}
			\end{figure}

Next, we focus on the error rates of the BBM92 protocol using $\rho^\perp$ as the EPR pair. The bit and phase error rates for such a protocol are given as,
\begin{eqnarray}
&& e_b=\text{Tr}\left(P_0^{\text{A}} P_0^{\text{B}} \Lambda^{\text{AD}}(\rho^{\perp}_{AB})\right) + \text{Tr} \left(P_1^{\text{A}} P_1^{\text{B}} \Lambda^{\text{AD}}(\rho^{\perp}_{AB})\right), \nonumber \\ 
&& e_p=\text{Tr} \left(P_+^{\text{A}} P_+^{\text{B}} \Lambda^{\text{AD}}(\rho^{\perp}_{AB})\right) + \text{Tr} \left(P_-^{\text{A}} P_-^{\text{B}} \Lambda^{\text{AD}}(\rho^{\perp}_{AB})\right),
\label{eq:error rate bbm92 perp}
\end{eqnarray}
where $\Lambda()^{\text{AD}}$ represents the action of the AD channel on the anti-correlated EPR pair. Again,  Charlie generates the EPR pair and the initial state is,
\begin{equation}
   \rho_{\text{Charlie}}^{\perp}=\frac{1}{2}\begin{pmatrix}
0 & 0 & 0 & 0\\
0 & 1 & -1 & 0 \\
0 & -1 & 1 & 0 \\
0 & 0 & 0 & 0
\end{pmatrix}. 
\end{equation}
When Charlie sends one-half of the EPR pair to Alice and Bob, the AD channel again acts independently on each half of the EPR pair, resulting in the final density matrix, 
\begin{equation}
   \rho_{\text{AB}}^{\perp}=\frac{1}{2}\begin{pmatrix}
2\gamma & 0 & 0 & 0\\
0 & 1-\gamma & \gamma-1 & 0 \\
0 & \gamma-1 & 1-\gamma & 0 \\
0 & 0 & 0 & 0
\end{pmatrix}. 
\end{equation}
We thus obtain the bit and phase error rates for the BBM92 protocol using $\rho^{\perp}$, as,
\begin{equation}
e_b=\gamma, \quad e_p=\frac{\gamma}{2}.
\end{equation}
Fig.~\ref{fig:theory_EPR} compares the performance of the BBM92 QKD protocol using the correlated and the anti-correlated EPR pairs, respectively. We observe that the protocol based on the correlated EPR pair offers a marginal advantage over the protocol based on an anti-correlated EPR pair, in the presence of an AD noise.

\subsubsection{Optimal strategy for distribution of EPR pairs under AD noise:}
Key generation in BBM92 relies on Alice and Bob sharing one half of an EPR pair each. This can be achieved by Alice generating an EPR pair and sending one half of the pair to Bob, or, by a third party (Charlie) generating an EPR pair and sending one half to Alice and Bob each. We now address the question of which of the two is better suited for the BBM92 protocol, assuming that the EPR pairs are shared over a noisy quantum channel.

Let us first consider the scenario where a third party, Charlie, generates a maximally correlated Bell state and sends one half to Alice and Bob each. We assume that Charlie is sitting precisely between Alice and Bob's lab. This scenario results in both Alice and Bob's qubits being affected by AD noise. However, the qubits have to traverse a shorter channel length compared to the scenario where Alice sends the qubits to Bob's lab. Hence, the damping probability for both Alice's and Bob's qubits is lower. 

\begin{figure}[h!]
				\centerline{\includegraphics[width=0.4\textwidth,height=\textheight,keepaspectratio]{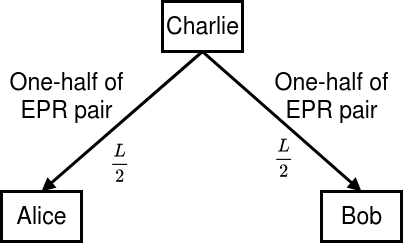}}
				\caption{Charlie sends one-half of an EPR pair to both Alice and Bob over an amplitude damped channel of length $\frac{L}{2}$.}
				\label{fig:bbm92 setup}
			\end{figure}
In the other scenario, Alice generates a state of form $\ket{\psi}_{AB}=\frac{1}{\sqrt{2}}\left(\ket{00}_{AB}+\ket{11}_{AB}\right)$. She sends one half of this maximally correlated Bell state to Bob using an AD channel. In this scenario, only Bob's half of the EPR pair gets exposed to the damping noise. However, Bob's qubit remains exposed for a longer duration compared to the scenario where Charlie sends the qubits to Alice and Bob both.

Recall that the damping probability $\gamma(t)$ is a function the time $t$ a qubit is exposed to the AD noise. In the first scenario, both Alice and Bob's qubits are exposed to AD noise for time $t$ as they traverse a channel length of $\frac{L}{2}$ each. The damping probability grows with $t$ as \cite{nielsen2002quantum},
\begin{equation}
\gamma(t) = 1-e^{-\frac{t}{T1}}, \label{eq:exp_gamma}
\end{equation}
where $T_1$ is the qubit relaxation time. Alternately, when Alice generates the EPR pair and sends one half to Bob, the qubit has to traverse a channel length of $L$. Hence, it remains exposed to the AD noise for time $2t$. Therefore, the damping probability is now,
\begin{equation}
  \gamma'(t)=1-e^{-\frac{2t}{T1}}.  
\end{equation}
Henceforth, we drop the dependence of $\gamma$ on time for notational simplicity. Taking logarithm followed by algebraic manipulation gives
\begin{equation}
    \gamma=2\gamma-\gamma^2.
\end{equation}

\begin{figure}[h!]
				\centerline{\includegraphics[width=0.4\textwidth,height=\textheight,keepaspectratio]{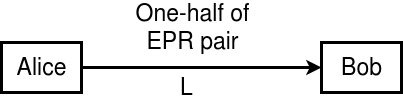}}
				\caption{Alice sends half of an EPR pair to Bob over a noisy channel of length $L$}
				\label{fig:bb84 setup}
			\end{figure}
In the second scenario where Alice shares one-half of her EPR pair with Bob,  her initial state ($\rho_{\text{ini}}$) is of the form shown in Eq.~\eqref{eq:cor den mat}. When she transmits one-half of EPR pair to Bob, AD noise acts on Bob's qubit as, 
\begin{eqnarray}
\rho_{AB}&=&\frac{1}{2}\sum_{i=0}^{i=1}\Big(I\otimes A_i^{\text{AD}}\left(\rho_{\text{ini}}\right) I\otimes A_i^{\text{AD}\dagger}\Big).
\end{eqnarray}
The resultant density matrix is
\begin{equation}
   \rho_{\text{AB}}^{\vert\vert}=\frac{1}{2}\begin{pmatrix}
1 & 0 & 0 & \sqrt{1-\gamma'}\\
0 & 0 & 0 & 0 \\
0 & 0 & \gamma' & 0 \\
\sqrt{1-\gamma'} & 0 & 0 & 1-\gamma'
\end{pmatrix}.
\end{equation}
We thus obtain the error rates of the BBM92 protcol when Alice prepares and distributes the EPR pair as,
\begin{equation}
e_b=\gamma-\frac{\gamma^2}{2},\quad e_p=\frac{\gamma}{2}.
\end{equation}

\begin{figure}[htbp!]
				\centerline{\includegraphics[width=0.5\textwidth,height=\textheight,keepaspectratio]{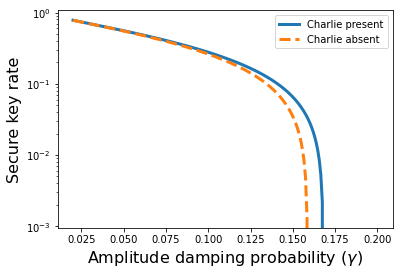}}
				\caption{Comparison of secure key rates of BBM92 QKD protocol for two different distribution schemes of EPR pairs.}
				\label{fig:EPR_dist}
			\end{figure}
			
Fig.~\ref{fig:EPR_dist} compares the performance of BBM92 for two scenarios, namely, when Charlie distributes the EPR pair to Alice and Bob and when Alice directly sends one-half of EPR pair to Bob. We observe that the secret kay rate shows marginal improvement when Charlie distributes the EPR pair in the presence of AD noise.

\subsubsection{Effect of channel asymmetry on BBM92:}
As shown in the previous subsection, the optimal strategy for sharing an EPR pair between Alice and Bob for implementing the BBM92 protocol is when a third party generates an EPR pair and sends one-half of the pair to Alice and Bob each. The channel between Charlie and Alice, and Charlie and Bob may have different lengths, leading to different damping probabilities. 
Let $\gamma_A$ and $\gamma_B$ be the damping probability of the AD channel between Charlie and Alice, and Charlie and Bob, respectively. 


As before, we assume that a third party, Charlie, generates the EPR pair $\rho^{\parallel}$ in Eq.~\eqref{eq:cor den mat}.
Recall that when Charlie sends one-half of the EPR pair over a noisy channel to Alice and Bob, 
the resultant density matrix is,
\begin{equation}
\rho_{\text{AB}}^{\parallel}=\frac{1}{2}\begin{pmatrix}
1+\gamma_A\gamma_B & 0 & 0 & \sqrt{(1-\gamma_A)(1-\gamma_B)}\\
0 & \gamma_A(1-\gamma_B) & 0 & 0 \\
0 & 0 & (1-\gamma_A)\gamma_B & 0 \\
\sqrt{(1-\gamma_A)(1-\gamma_B)} & 0 & 0 & (1-\gamma_A)(1-\gamma_B)
\end{pmatrix}. 
\end{equation}
Using Eq.~\eqref{eq:error rate bbm92}, we find that the QBER of the asymmetric BBM92 protocol using $\rho^{\parallel}_{AB}$ as the EPR pair is,
\begin{equation}
    e_b=\frac{1}{2}\left[\gamma_A(1-\gamma_B)+\gamma_B(1-\gamma_A)\right].
\end{equation}
Similarly, the phase error rate of the asymmetric BBM92 protocol using $\rho^{\parallel}_{AB}$ is
\begin{equation}
    e_p=\frac{1-\sqrt{(1-\gamma_A)(1-\gamma_B)}}{2}.
\end{equation}

\begin{figure}[h!]
				\centerline{\includegraphics[width=0.5\textwidth,height=\textheight,keepaspectratio]{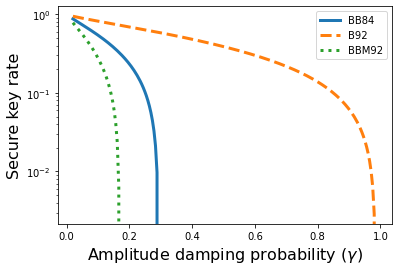}}
				\caption{Comparison of secure key rates of QKD protocols under AD noise.}
				\label{fig:theory_rates}
			\end{figure}

 Fig.~\ref{fig:theory_rates} shows the performance of different QKD protocols in the presence of AD noise. We have ignored system imperfections such as qubit read-out errors while obtaining the plots. We observe that the B92 protocol outperforms the other two protocols (BB84 and BBM92) by a wide margin. The superior performance of B92 under AD noise is due to the fact that $\ket{0}$ remains unaffected due to AD noise. Since half of the qubits in B92 is $\ket{0}$, thereby making it a suitable candidate for implementation when the channel is amplitude damped.
			
\section{Dual-rail encoded BB84}\label{sec:dual-rail}

Having quantified the effects of AD and GAD noise on QKD protocols, we now address the question of whether quantum error correction (QEC) can improve the secure key rates of such noisy QKD protocols.  
Rather than use a general-purpose stabilizer code~\cite{terhal2015} which maybe resource-intensive, we use a noise-adapted error mitigation procedure tailored for amplitude-damping noise. In particular, we  design an \emph{encoded} BB84 protocol based on the dual-rail qubits discussed in Sec.~\ref{subsec:dual-rail}, which are known to be robust against amplitude-damping. We present two different schemes for implementing dual-rail encoded BB84 with error detection and post-selection. We also compare the secure key rates for these different QEC-based protocols.

\subsection{Dual-rail encoded BB84 with ancilla-based post-selection technique}\label{sec: PS with ancilla}

\begin{center}
    \begin{figure}[h]
    \centerline{\includegraphics[width=0.9\textwidth,height=\textheight,keepaspectratio]{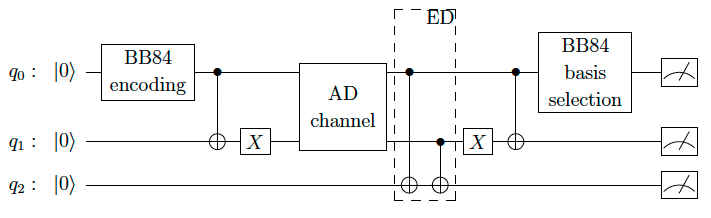}}
\caption{Dual-rail encoded BB84 with post-selection technique. ED is the Error Detection block, comprising of 2 \textsc{cnot}s.}
				\label{fig:dual bb84}
			\end{figure}
\end{center}

In the encoded BB84 protocol, Alice encodes her states $\{\ket{0}, \ket{1}, \ket{+}, \ket{-}\}$ using the dual-rail encoding described in Eq.~\eqref{eq:dual}. She sends her encoded qubits to Bob through an amplitude-damping channel. Bob uses an ancilla qubit and a pair of consecutive \textsc{cnot} gates to detect the damping errors, as shown in Fig.~\ref{fig:dual bb84}. We have shown in  Sec.~\ref{subsec:dual-rail} that Bob's error detection strategy can detect the occurrence of amplitude-damping for any arbitrary qubit state (see Fig.~\ref{fig:detection}). If an AD error occurred, that is, if Bob obtains $\ket{0}$ after the ancilla measurement, he throws away the received encoded state. On the other hand, if Bob does not detect any AD error, he uses a decoder circuit to get back the BB84 state. Finally, as in the final step of the standard BB84 protocol, Bob measures his decoded BB84 states in either the $X$ or $Z$ basis, to obtain his raw key bits. We refer to this process of discarding AD errors and using only using error-free qubits to get the raw key as the \emph{post-selection} (PS) technique. After obtaining the raw key, Alice and Bob do classical post-processing to arrive at the shared secret key.

The protocol described above detects the occurrence of amplitude damping errors with absolute certainty, so the bit and phase error rates due to damping errors shown in Eqs.~\eqref{eq:bit error rate bb84} and~\eqref{eq:phase error rate bb84} vanish altogether. However, the post-selection step employed in our dual-rail BB84 discards a fraction of transmitted key bits, making the sifting factor equal to $1-\gamma$. Note that we are only concerned with the error rates due to AD noise and do not consider the effects of an eavesdropper or any other channel noise on our dual-rail encoded protocol.
\subsubsection{Effect of imperfect \textsc{cnot}s on the performance of dual-rail encoded BB84:}
In our previous analysis, we have assumed that the encoding and decoding circuits are built from perfect quantum gates. However, the NISQ devices of today do not have gates with perfect fidelity. In this subsection, we study the effects of \textsc{cnot} gate imperfection on the secure key rate of the dual-rail BB84 QKD protocol. In our study, we do not assume that all the \textsc{cnot}s in the circuit fail at once. Rather, we draw inspiration from the theory of fault-tolerant quantum computing~\cite{preskill98}, where a critical sub-circuit is assumed to be noisy while the remaining circuit works perfectly. Along similar lines, we look at how the \textsc{cnot} failure in the encoder circuit, post-selection circuit, and decoder circuit, separately affect the performance of the BB84 protocol. We model the \textsc{cnot} gate operation as ~\cite{briegel1998quantum},
\begin{equation}
    \rho_{\text{out}}=(1-\beta) \textsc{cnot}_{i\to j}  \left(\rho_{\text{in}}\right) {\textsc{cnot}^{\dagger}}_{i\to j}+\frac{\beta}{4} I_{i,j},
\label{eq:CNOT imperfect}
\end{equation}
where $\rho_{\text{in}}$ ($\rho_{\text{out}}$) is the two-qubit input (output) state before (after) \textsc{cnot},  $\beta$ is the failure probability of the \textsc{cnot} gate and $I_{i,j}$ is the two-qubit identity operator. $\textsc{cnot}_{i\to j}$ represents the unitary matrix for an ideal \textsc{cnot} operation with $i$ as the control and $j$ as the target qubit. Eq.~\eqref{eq:CNOT imperfect} models the action of a noisy \textsc{cnot} as a mixture of the action of an ideal \textsc{cnot} with  probability $1-\beta$ and the action of a depolarizing channel with noise parameter $\beta$ on qubits $i$ and $j$.

\begin{figure}[h!]
				\centerline{\includegraphics[width=0.5\textwidth,height=\textheight,keepaspectratio]{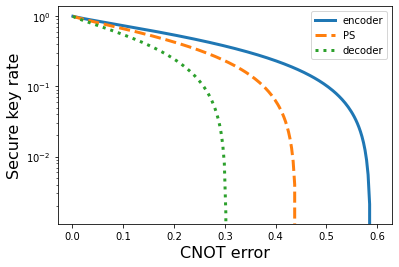}}
				\caption{Effect of imperfect \textsc{cnot} on dual-rail encoded BB84 with ancilla-based PS technique.}
				\label{fig:cnot_error}
			\end{figure}

Table~\ref{table:imperfect cnot} shows the effect of imperfect \textsc{cnot} gates on the bit and phase error rates of the dual-rail encoded BB84 protocol. 
\begin{table}
\caption{\label{table:imperfect cnot}Effect of imperfect \textsc{cnot} in different subcircuits on $e_b$ and $e_p$.}
\begin{tabular*}{\textwidth}{@{}l*{15}{@{\extracolsep{0pt plus
12pt}}l}}
\br
 &Bit error rate&Phase error rate\\
\mr
Imperfect encoder & $\frac{\beta}{4}(1-\gamma)(1+\gamma)$ &$\frac{\beta}{4}(1-\gamma)(1+\gamma)$\\ \\
Imperfect post-selection & $\frac{\beta}{4}$&$\frac{\beta}{4}$\\ \\
Imperfect decoder & $\frac{\beta}{2}(1-\gamma)$ & $\beta(1-\gamma)$\\
\br
\end{tabular*}
\end{table}
Note that we have assumed the post-selection subcircuit is imperfect due to the noisy  $\textsc{cnot}_{q_0\to q_2}$ in Fig.~\ref{fig:dual bb84}. Fig.~\ref{fig:cnot_error} shows how the \textsc{cnot} imperfections in different sub-circuits affect the performance of the overall schematic that implements dual-rail encoded BB84 with the PS technique. We have assumed the damping parameter  $\gamma$ to be equal to $0.5$, a value at which conventional (unencoded) BB84 does not produce any secure key rate, while obtaining the plots shown in Fig~\ref{fig:cnot_error}. We observe that the decoder \textsc{cnot} error has the highest impact on the final secure key rate. The encoder and post-selection \textsc{cnot}s have a lesser impact, since some of the erroneous states get discarded during post-selection.

\subsection{Dual-rail encoded BB84 with optimal post-selection technique}\label{subsec:optimal PS}

As described in the previous subsection, \textsc{cnot} imperfections deteriorate the performance of QKD protocols. The ancilla-based post-selection technique requires a decoder, two \textsc{cnot}s in the post-selection subcircuit, and an additional ancilla qubit. Hence, we use an optimal post-selection circuit with minimal resources for improving the performance of BB84 QKD in the presence of AD noise. Such a post-selection circuit has been used in dual-rail state transfer protocols in the past~\cite{bose2005}.

\begin{figure}[h]
    \centerline{\includegraphics[width=0.9\textwidth,height=\textheight,keepaspectratio]{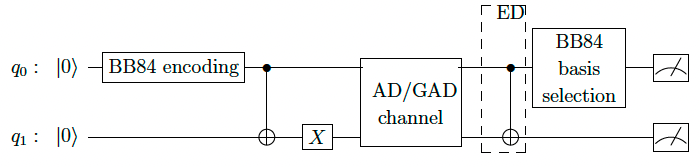}}
\caption{Optimal dual-rail encoding with post-selection. ED is the Error Detection block, comprising of 2 \textsc{cnot}s.}
				\label{fig:optimal dual bb84}
			\end{figure}

Fig.~\ref{fig:optimal dual bb84} shows the \emph{optimal} post-selection circuit, which requires only an encoder \textsc{cnot} and a post-selection \textsc{cnot}. In this method, Bob uses the second qubit of the dual-rail encoding for post selection. If he measures the second qubit as $\ket{0}$, he knows that an amplitude-damping error has occurred and discards that particular key instance. We have shown the efficacy of this circuit in detecting amplitude damping of an arbitrary state, $\alpha\ket{0}+\beta\ket{1}$, in \ref{Appendix: arbitrary state optimal PS}. Similar to the earlier post-selection technique, the optimal post-selection technique also results in zero bit and phase error rates. The sifting factor of this optimal post-selection circuit has a dependence on $\gamma$ identical to that of ancilla-based post-selection technique.

\begin{figure}[h!]
				\centerline{\includegraphics[width=0.5\textwidth,height=\textheight,keepaspectratio]{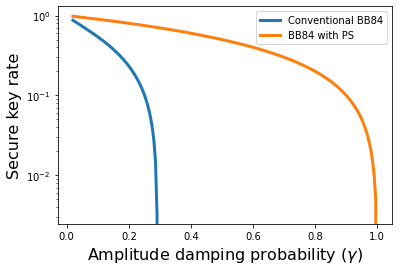}}
				\caption{Performance of dual-rail encoded BB84 under amplitude-damping noise.}
				\label{fig:theory_bb84_variants}
			\end{figure}

Fig.~\ref{fig:theory_bb84_variants} compares the performance of conventional BB84 and dual-rail BB84 with post-selection technique under AD noise. We have not taken any system imperfections while obtaining the secure key rates plots. We again emphasize that dual-rail BB84 with ancilla-based post-selection and dual-rail BB84 with optimal post-selection circuit behaves identically in the absence of any system imperfections, and both give a secure key fraction of $1-\gamma$. We observe that BB84 with post-selection technique outperforms the conventional BB84 by a wide margin.  

\begin{figure}[h!]
				\centerline{\includegraphics[width=0.7\textwidth,height=\textheight,keepaspectratio]{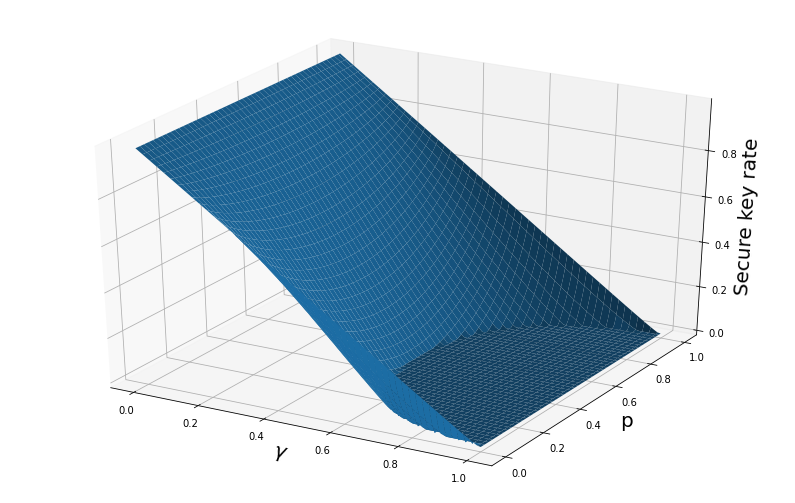}}
				\caption{Secure key rate of dual-rail encoded BB84 with post-selection technique in the presence of GAD noise.}
				\label{fig:GAD}
			\end{figure}

Finally, we point out that both the post-selection circuits (optimal circuit and ancilla-based circuit) are capable of detecting a fraction of errors caused by the generalized amplitude-damping channel and hence can provide a nonzero secure key rate for the BB84 protocol in the presence of GAD noise. We have shown in ~\ref{sec:GAD_dual} that GAD noise maps the logical $|0\rangle_{L}$ to one of four possible states - $\ket{01}$, $\ket{10}$, $\ket{00}$ and $\ket{11}$. The \textsc{cnot} gates of the post-selection circuit can ascertain the parity of the dual-rail qubits and hence detect errors due to the mapping of the BB84 logical qubits to $\ket{00}$ and $\ket{11}$. However, our post-selection circuits cannot detect the two-qubit GAD errors, thereby introducing bit and phase errors in the protocol (see Eq.~\eqref{eq:GAD_error_dual}). Fig.~\ref{fig:GAD} shows the dependence of the secure key rate for the dual-rail encoded BB84 protocol with post-selection, on both the noise parameters $\gamma$ and $p$ of the GAD channel.

\section{Implementation on IBM quantum computers}\label{sec:Implementation}

We now describe how the dual-rail BB84 protocols described in Sec.~\ref{sec:dual-rail}
can be tested by implementing them on NISQ devices. Our implementation makes use of the superconducting qubits on the IBM quantum processors, for which amplitude-damping noise constitutes a natural source of noise \cite{etxezarreta2021time,kubica2022erasure}. In fact, we assume that AD noise is one of the dominant sources of noise on a superconducting quantum processor, an assumption that is justified in the following section.

\subsection{Implementing amplitude-damping noise on IBMQ}

\begin{figure}[h]
\centerline{\includegraphics[width=0.7\textwidth,height=\textheight,keepaspectratio]{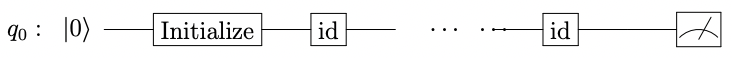}}
    \caption{Circuit with a train of identity gates to mimic the AD channel.}
				\label{fig:damp}
			\end{figure}
			
One approach to implementing the amplitude-damping channel on the IBM processors is to pass a qubit through a chain of identity gates before measuring it, as shown in Fig.~\ref{fig:damp}. Noise acts on the qubit for the duration for which the it remains idle. We first verify that AD noise is indeed one of the dominant sources of noise on idling qubits in a superconducting quantum processor. 	

\begin{figure}
     \centering
     \begin{subfigure}[b]{0.3\textwidth}
         \centering
         \includegraphics[width=\textwidth]{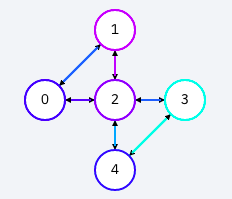}
         \caption{}
        \label{subfig:damp}
     \end{subfigure}
     \begin{subfigure}[t]{0.4\textwidth}
         \centering
         \includegraphics[width=\textwidth]{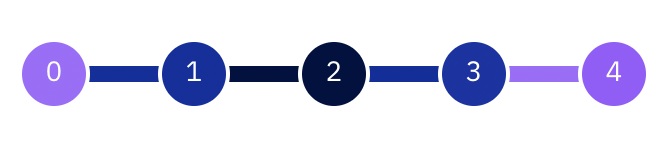}
         \caption{}
         \label{subfig:bogota}
     \end{subfigure}
        \caption{(a) IBM Yorktown connectivity (b) IBM Bogota connectivity.}
        \label{fig:connectivity}
\end{figure}

We implement the verification circuit in Fig.~\ref{fig:damp} on the IBM Yorktown processor\footnote{IBM Yorktown is an old generation quantum processor and has been decommissioned recently.}. 
The important properties of the qubits of IBM Yorktown are shown in Table~\ref{table:York}. We focus on on qubits $3$ and $4$ and initialize these two qubits as $\ket{01}$. The operation time of an identity gate in the IBM Yorktown processor is $35.6$ ns. 

\begin{figure}[h!]

				\centerline{\includegraphics[width=0.5\textwidth,height=0.8\textheight,keepaspectratio]{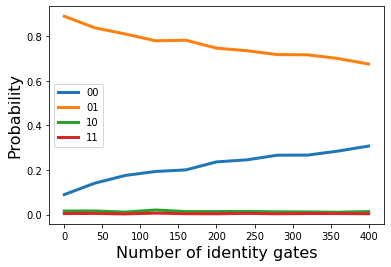}}
				\caption{Measurement results for different delays when the initial state is $\ket{01}$.}
				\label{fig:delay}
			\end{figure}

Fig.~\ref{fig:delay} shows the measurement statistics for different delays, with the qubits starting in the state $\ket{01}$. We first observe that the intrinsic gate imperfections and read-out errors lead to errors even in the absence of any delay. We further observe that when the qubits start in the $\ket{01}$ state, the probability of measuring $\ket{10}$ and $\ket{11}$ remains nearly constant as we increase the delay. This hints that noise models such as GAD and bit-flip noise are not as dominant in the IBM Yorktown processor. Finally, we see that the probability of $\ket{01}$ going to $\ket{00}$ increases as we increase the delay, thereby confirming our assumption that AD noise being one of the dominant noise models in the processor. Recall that the effect of AD noise must increase with an increase in the delay due to exponential dependence of $\gamma$ on time (see Eq.~\eqref{eq:exp_gamma}).  We note in passing that phase damping is indeed another dominant noise model for the IBM quantum processors. However, this work focuses only on mitigating the effects of AD noise.

\begin{table}
\caption{\label{table:York}Specifications of qubits of IBM Yorktown.}
\begin{tabular*}{\textwidth}{@{}l*{15}{@{\extracolsep{0pt plus
12pt}}l}}
\br
 Qubit & $T_1$ ($\mu$s) & Readout error ($\%$)\\
\mr
Qubit $0$ & $44.33$ &$10.7$\\
Qubit $1$ & $50.67$ &$35.6$\\
Qubit $2$ & $70.27$ &$7.9$\\
Qubit $3$ & $57.62$ &$3$\\ 
Qubit $4$ & $56.94$ &$5.4$\\
\br
\end{tabular*}
\end{table}

 \subsection{Implementation of BB84 QKD protocol on IBMQ}
 
 We next describe our implementation of the BB84 protocol on the IBM Yorktown processor. This processor has large qubit read-out errors, and hence is an excellent testbed for studying the effects of read-out errors on the performance of the BB84 protocol. As shown in Fig.~\ref{fig:damp}, we use a train of identity gates to mimic the amplitude-damping channel. Bob, who sits at the same physical qubit as Alice, randomly measures the received states in the $X$ or $Z$ basis and publicly announces the measurement result. Alice and Bob extract a shared bit string using a public classical channel. Fig.~\ref{fig:bb84 setup} shows the BB84 QKD schematic used in our implementation. 
 \begin{figure}[h!]
				\centerline{\includegraphics[width=0.6\textwidth,height=\textheight,keepaspectratio]{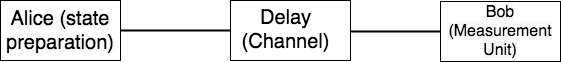}}
				\caption{Schematic of the BB84 setup. Identity gates introduce delay.}
				\label{fig:bb84 setup}
			\end{figure}

Our implementation proceeds as follows. Alice `sends' $\ket{0}$ in a block of $8192$ qubits, followed by a block of $\ket{1}$, $\ket{+}$, and $\ket{-}$. Note that, we use $8192$ as the block length because IBM processors limit the number of shots in a run to $8192$. Hence, we have a sifted key length of $32,768$. As we are interested in the effects of damping on the BB84 states in our study, we do not worry about eavesdropping due to lack of randomness in our sifted key. This block-wise transmission of the BB84 states reduces our quantum processor access time drastically. Based on the classical postprocessing, we estimate the quantum bit error rate (QBER) and obtain the secure key rate using Eq.~\eqref{eq:key}.

\begin{figure}[h!]
				\centerline{\includegraphics[width=0.5\textwidth,height=\textheight,keepaspectratio]{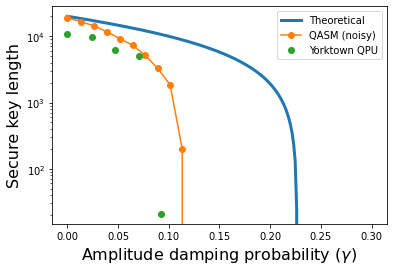}}
				\caption{BB84 Secure key rate with $3\%$ as qubit read-out error.}
				\label{fig:bb84_3percent}
			\end{figure}

Fig.~\ref{fig:bb84_3percent} shows the dependence of secure key length and QBER on the damping probability for qubit $3$ of IBM Yorktown processor. We also implement this setup on the QASM simulator, where we import the noise model and connectivity of the Yorktown processor in our simulations.

We observe that the QBER increases with an increase in the damping probability. This increase in QBER results in the reduction of secure key length.

\subsection{Implementation of  BBM92 QKD protocol on IBMQ}
We implement the BBM92 QKD protocol on qubits $0$ and $1$ of IBM Bogota (see Fig.~\ref{subfig:bogota}). The BBM92 implementation assumes that a third party, Charlie distributes the EPR pairs to Alice and Bob (see Fig.~\ref{fig:bbm92 setup}). As before, the amplitude-damping channels between Alice and Charlie, and, Bob and Charlie are both realised by a train of identity gates.
\begin{table}\label{Table:bogota}
\caption{Specifications of qubits of IBM Bogota.}
\begin{tabular*}{\textwidth}{@{}l*{15}{@{\extracolsep{0pt plus
12pt}}l}}
\br
 Qubit & $T_1$ ($\mu$s) & Readout error ($\%$)\\
\mr
Qubit $0$ & $97.6$ &$3.2$\\
Qubit $1$ & $218.2$ &$1.94$\\
Qubit $2$ & $200.3$ &$6.03$\\
Qubit $3$ & $111.3$ &$5$\\ 
Qubit $4$ & $151.1$ &$1.78$\\
\br
\end{tabular*}
\end{table}
 
\begin{figure}[h!]
				\centerline{\includegraphics[width=0.5\textwidth,height=\textheight,keepaspectratio]{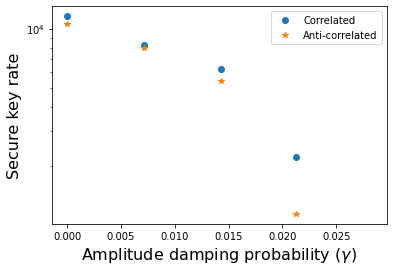}}
				\caption{BBM92 with correlated and anti-correlated EPR pairs on IBM Bogota}
				\label{fig:exp_bbm92}
			\end{figure}
			
Fig.~\ref{fig:exp_bbm92} shows that correlated EPR-based BBM92 protocol performs slightly better than anti-correlated EPR based BBM92, thus confirming our analytical discussion in Sec.~\ref{subsub:EPR pairs}. We also observe that the key rate drops to zero after a damping probability of $0.02$ because of qubit readout errors and \textsc{cnot} imperfections. 

\subsection{Implementation of B92 QKD protocol on IBMQ}

\begin{figure}[h!]
				\centerline{\includegraphics[width=0.5\textwidth,height=\textheight,keepaspectratio]{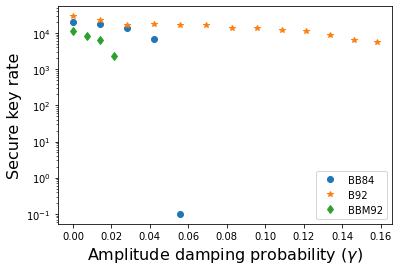}}
				\caption{Secure key rates of QKD protocols implemented on IBM Bogota.}
				\label{fig:exp_key}
			\end{figure}

Similar to the BB84 implementation in Fig.~\ref{fig:bb84 setup}, we implement B92 QKD on qubit $0$ of IBM Bogota. Fig.~\ref{fig:exp_key} compares the performance of BB84, B92 and BBM92 implementations on IBM Bogota. We have used qubit $0$ of IBM Bogota to implement BB84 and B92 protocols and qubits $0$ and $1$ for implementing the BBM92 protocol. As per our expectations (see Fig.~\ref{fig:theory_rates}), we observe that the B92 protocol outperforms the other two protocols (BB84 and BBM92) in our experimental implementations too.

\subsection{Implementation of dual-rail encoded BB84 protocol on IBMQ}

Finally, we present the results of our implementation of the two different encoded BB84 schemes (ancilla-based and optimal post-selection techniques) on the IBM quantum processors. We use IBM Yorktown and IBM Bogota for our implementations.

\subsubsection{Implementation of dual-rail encoded BB84 protocol using ancilla-based post-selection technique}

We implement the schematic shown in Fig.~\ref{fig:dual bb84} on IBM Yorktown. In Fig.~\ref{fig:dual bb84}, the \textsc{cnot}s in encoder and post-selection subcircuits are oriented such that the three qubits of the circuit need to be on a triangular lattice. IBM Yorktown offers such a triangular connectivity, thereby removing the requirement of any \textsc{swap} gate. However, high \textsc{cnot} gate errors and qubit readout errors make the intrinsic QBER, i.e, QBER in the absence of any AD, more than $11\%$, leading to a zero secret key rate. Hence, we implement our ancilla-based post-selection technique on the QASM simulator with the connectivity and noise metrics mimicking the IBM Yorktown machine.

\begin{figure}[h!]
				\centerline{\includegraphics[width=0.5\textwidth,height=\textheight,keepaspectratio]{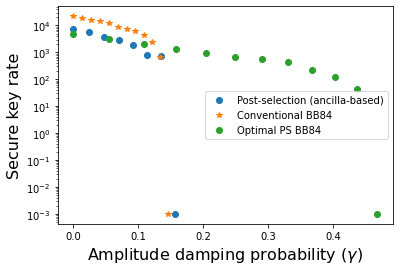}}
				\caption{Secure key rates of dual-rail encoded BB84 on noisy QASM simulator.}
				\label{fig:sim_optimal_PS}
			\end{figure}

Fig.~\ref{fig:sim_optimal_PS} compares the performance BB84, dual-rail BB84 with ancilla-based post-selection technique and dual-rail BB84 with optimal post-selection technique on a noisy QASM simulator. Dual-rail encoded protocol with ancilla-based post-selection technique performs better than BB84 protocol in theory, but the advantages of the protocol get subdued due to the presence of a number of \textsc{cnot} gates and results being dependent on the correct readout of three qubits. Hence, our implementation shows the need to reduce the number of qubits and \textsc{cnot}s in the circuit. We have obtained such a reduction in the circuit complexity using the schematic shown in Fig.~\ref{fig:optimal dual bb84},resulting in better performance than conventional BB84 on the QASM simulator, as evident from Fig.~\ref{fig:sim_optimal_PS}. 

\subsubsection{Implementation of dual-rail encoded BB84 protocol using optimal post-selection technique}
\begin{figure}[h!]
				\centerline{\includegraphics[width=0.45\textwidth,height=\textheight,keepaspectratio]{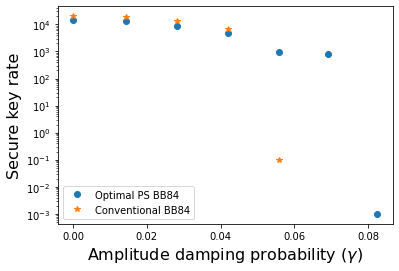}}
				\caption{Secure key rates of dual-rail encoded BB84 using an optimal post-selection technique on IBM Bogota processor.}
				\label{fig:exp_optimal_PS}
			\end{figure}

We implement the schematic shown in Fig.~\ref{fig:optimal dual bb84} on IBM Bogota. The optimal post-selection technique does not require a triangular lattice and hence allows us to use the new generation of IBM processors with better \textsc{cnot} fidelity and lower read-out errors. Dual-rail encoded protocol performs better compared to BB84 protocol by a wide margin. Fig.~\ref{fig:exp_optimal_PS} compares the performance of BB84 and dual-rail BB84 using optimal post-selection technique on IBM Bogota. We observe that the dual-rail encoded protocol can withstand higher AD noise compared to the conventional BB84 implementation on a superconducting quantum processor.

\section{Summary and Outlook}\label{sec:Conclusion}

We have studied the effects of AD noise on the secure key rate of different QKD protocols both analytically and using a noisy quantum processor. Specifically, in the presence of amplitude-damping noise, we have characterized the effect of channel asymmetry, the distribution scheme  and type of entangled qubit pairs, on the secure key rate of an entanglement-based protocol like BBM92. We have demonstrated the superior performance of B92 over BB84 and BBM92 QKD protocols in the presence of AD noise. We have implemented the BB84 QKD protocol using dual-rail encoding with a post-selection technique and shown the pros and cons of such an encoding over the conventional BB84 protocol in the presence of AD noise. The effects of gate imperfections and qubit read-out errors on the performance of various QKD protocols have also been quantified. 

Our specific choice of noise model -- the amplitude-damping noise channel -- is motivated by several factors. The damping noise parameter has an exponential dependence on channel delay. In parallel, fiber-based QKD implementations focus on the probability of photon loss in a channel, which is again exponentially dependent on channel length. Hence, our QKD implementation over an amplitude damping channel using a QC shows similar variations in secure key rate as fiber-based implementations. As another point of comparison, fiber-based implementations of BB84 typically use the polarization degree of freedom of a photon, in conjunction with a successful detection event as a dual-rail qubit. The loss of the polarized photon leads to AD error detection, whereas in our implementation, we use the parity of the codewords for AD error detection. In fact, dual-rail qubits effectively map damping errors into the more easily detectable erasure errors, as has been formally shown in the literature recently~\cite{kubica2022erasure}. 

We may note here that the damping probability in a quantum computer is higher than in a typical optical fiber-based QKD implementation. But this gap narrows down in the context of QKD protocols using quantum memories, which have been proposed as a means to increase the transmission distance \cite{panayi2014memory,piparo2017memory}. Quantum memories are prone to damping noise, thereby increasing the chances of damping errors in typical fiber-based QKD implementations as well~\cite{hetet2008quantum,otten2021impacts}. Hence, the tools presented in our study would be suited for detecting AD errors in memory-assisted QKD protocols. Note that, while our work captures the contribution of gate imperfections and measurement error on the secure key rate, a detailed modelling of optical channel and detectors using a QC is still required to mimic practical QKD implementations over optical fiber networks.

Our work is also a first step towards understanding how quantum error correction can impact the secure key rates of different QKD protocols, when they are implemented over noisy channels. Moving forward we would like to study how such error mitigation techniques can enable long-distance QKD using quantum repeaters realised via noisy quantum memories. 

Finally, this study assumes that only a single \textsc{cnot} failure occurs in the entire circuit while estimating the effects of \textsc{cnot} failure on the secure key rate. One future direction would be to develop efficient numerical techniques that can show the effects of a series of \textsc{cnot} gate failures on the performance of QKD protocols. Recent works have focused on developing such numerical techniques for finding the secure key rate of a QKD protocol using a quantum repeater with encoding \cite{Jing_2020,jing2022quantum}. We also point out that our QKD implementation generates finite length key bits, and the size of these finite key blocks is small (of the order of $10^3$). Hence, the key generated from our implementation can be used to verify the recent theoretical works on the security of QKD protocols with small block length \cite{lim2021}.
\section*{Acknowledgements}
We acknowledge the use of IBM Quantum for this work. The views expressed are those of the authors and do not reflect the official policy or position of IBM or the IBM Quantum team. This research was supported in part by a grant from the Mphasis F1 Foundation to the Centre for Quantum Information, Communication, and Computing (CQuICC). PM acknowledges financial support from the Department of Science and Technology, Government of India, under Grant No. DST/ICPS/QuST/Theme-3/2019/Q59.
\section*{References}
\bibliographystyle{unsrt}
\bibliography{ref}

\begin{thebibliography}{10}

\bibitem{bennett2020quantum}
Charles~H. Bennett and Gilles Brassard.
\newblock {Quantum cryptography: Public key distribution and coin tossing}.
\newblock {\em Theoretical Computer Science}, 560:7--11, Dec 2014.

\bibitem{gisin2002quantum}
Nicolas Gisin, Gr{\'e}goire Ribordy, Wolfgang Tittel, and Hugo Zbinden.
\newblock {Quantum cryptography}.
\newblock {\em Reviews of Modern Physics}, 74(1):145, 2002.

\bibitem{pirandola2019advances}
Stefano Pirandola, Ulrik~L Andersen, Leonardo Banchi, Mario Berta, Darius
  Bunandar, Roger Colbeck, Dirk Englund, Tobias Gehring, Cosmo Lupo, Carlo
  Ottaviani, et~al.
\newblock {Advances in quantum cryptography}.
\newblock {\em Advances in Optics and Photonics}, 12(4):1012--1236, 2020.

\bibitem{Xu_2020}
Feihu Xu, Xiongfeng Ma, Qiang Zhang, Hoi-Kwong Lo, and Jian-Wei Pan.
\newblock {Secure quantum key distribution with realistic devices}.
\newblock {\em Reviews of Modern Physics}, 92(2), May 2020.

\bibitem{liao2017satellite}
Sheng-Kai Liao, Wen-Qi Cai, Wei-Yue Liu, Liang Zhang, Yang Li, Ji-Gang Ren,
  Juan Yin, Qi~Shen, Yuan Cao, Zheng-Ping Li, et~al.
\newblock {Satellite-to-ground quantum key distribution}.
\newblock {\em Nature}, 549(7670):43--47, 2017.

\bibitem{feng2021experimental}
Zhao Feng, Shangbin Li, and Zhengyuan Xu.
\newblock {Experimental underwater quantum key distribution}.
\newblock {\em Optics Express}, 29(6):8725--8736, 2021.

\bibitem{arute2019quantum}
Frank Arute, Kunal Arya, Ryan Babbush, Dave Bacon, Joseph~C Bardin, Rami
  Barends, Rupak Biswas, Sergio Boixo, Fernando~GSL Brandao, David~A Buell,
  et~al.
\newblock {Quantum supremacy using a programmable superconducting processor}.
\newblock {\em Nature}, 574(7779):505--510, 2019.

\bibitem{Zhong1460}
Han-Sen Zhong, Hui Wang, Yu-Hao Deng, Ming-Cheng Chen, Li-Chao Peng, Yi-Han
  Luo, Jian Qin, Dian Wu, Xing Ding, Yi~Hu, et~al.
\newblock {Quantum computational advantage using photons}.
\newblock {\em Science}, 370(6523):1460--1463, 2020.

\bibitem{wu2021strong}
Yulin Wu, Wan-Su Bao, Sirui Cao, Fusheng Chen, Ming-Cheng Chen, Xiawei Chen,
  Tung-Hsun Chung, Hui Deng, Yajie Du, Daojin Fan, et~al.
\newblock {Strong quantum computational advantage using a superconducting
  quantum processor}.
\newblock {\em Physical Review Letters}, 127(18):180501, 2021.

\bibitem{preskill2018quantum}
John Preskill.
\newblock {Quantum computing in the NISQ era and beyond}.
\newblock {\em Quantum}, 2:79, 2018.

\bibitem{Watanabe_2008}
Shun Watanabe, Ryutaroh Matsumoto, and Tomohiko Uyematsu.
\newblock {Tomography increases key rates of quantum-key-distribution
  protocols}.
\newblock {\em Physical Review A}, 78(4):042316, 2008.

\bibitem{sano2010secure}
Yousuke Sano, Ryutaroh Matsumoto, and Tomohiko Uyematsu.
\newblock {Secure key rate of the BB84 protocol using finite sample bits}.
\newblock {\em Journal of Physics A: Mathematical and Theoretical},
  43(49):495302, 2010.

\bibitem{Som_teleportation2002}
Somshubhro Bandyopadhyay.
\newblock {Origin of noisy states whose teleportation fidelity can be enhanced
  through dissipation}.
\newblock {\em Physical Review A}, 65:022302, Jan 2002.

\bibitem{Jing_2020}
Yumang Jing, Daniel Alsina, and Mohsen Razavi.
\newblock Quantum key distribution over quantum repeaters with encoding:
  {U}sing error detection as an effective postselection tool.
\newblock {\em Physical Review Applied}, 14(6), Dec 2020.

\bibitem{jing2022quantum}
Yumang Jing and Mohsen Razavi.
\newblock Quantum repeaters with encoding on nitrogen-vacancy-center platforms.
\newblock {\em Physical Review Applied}, 18(2):024041, 2022.

\bibitem{jayashankar2022}
Akshaya Jayashankar and Prabha Mandayam.
\newblock Quantum error correction: Noise-adapted techniques and applications.
\newblock {\em Journal of the Indian Institute of Science}, pages 1--16, 2022.

\bibitem{nielsen2002quantum}
Michael~A. Nielsen and Isaac~L. Chuang.
\newblock {\em {Quantum Computation and Quantum Information: 10th Anniversary
  Edition}}.
\newblock Cambridge University Press, USA, 10th edition, 2011.

\bibitem{Zhukov_2018}
A.~A. Zhukov, E.~O. Kiktenko, A.~A. Elistratov, W.~V. Pogosov, and Yu.~E.
  Lozovik.
\newblock {Quantum communication protocols as a benchmark for programmable
  quantum computers}.
\newblock {\em Quantum Information Processing}, 18(1), Dec 2018.

\bibitem{bose2005}
Daniel Burgarth and Sougato Bose.
\newblock {Conclusive and arbitrarily perfect quantum-state transfer using
  parallel spin-chain channels}.
\newblock {\em Physical Review A}, 71:052315, May 2005.

\bibitem{PhysRevA.95.042342}
Wen-Jie Zou, Yu-Huai Li, Shu-Chao Wang, Yuan Cao, Ji-Gang Ren, Juan Yin,
  Cheng-Zhi Peng, Xiang-Bin Wang, and Jian-Wei Pan.
\newblock {Protecting entanglement from finite-temperature thermal noise via
  weak measurement and quantum measurement reversal}.
\newblock {\em Physical Review A}, 95:042342, Apr 2017.

\bibitem{chirolli2008decoherence}
Luca Chirolli and Guido Burkard.
\newblock {Decoherence in solid-state qubits}.
\newblock {\em Advances in Physics}, 57(3):225--285, 2008.

\bibitem{shor2000simple}
Peter~W Shor and John Preskill.
\newblock {Simple proof of security of the BB84 quantum key distribution
  protocol}.
\newblock {\em Physical Review Letters}, 85(2):441, 2000.

\bibitem{B921}
Charles~H. Bennett.
\newblock {Quantum cryptography using any two nonorthogonal states}.
\newblock {\em Physical Review Letters}, 68:3121--3124, May 1992.

\bibitem{E91}
Artur~K. Ekert.
\newblock {Quantum cryptography based on Bell's theorem}.
\newblock {\em Physical Review Letters}, 67:661--663, Aug 1991.

\bibitem{BBM92}
Charles~H. Bennett, Gilles Brassard, and N.~David Mermin.
\newblock {Quantum cryptography without Bell's theorem}.
\newblock {\em Physical Review Letter}, 68:557--559, Feb 1992.

\bibitem{erven2012experimental}
Chris Erven.
\newblock On experimental quantum communication and cryptography.
\newblock {\em University of Waterloo}, 2012.

\bibitem{terhal2015}
Barbara~M Terhal.
\newblock Quantum error correction for quantum memories.
\newblock {\em Reviews of Modern Physics}, 87(2):307, 2015.

\bibitem{preskill98}
John Preskill.
\newblock Fault-tolerant quantum computation.
\newblock In {\em Introduction to quantum computation and information}, pages
  213--269. World Scientific, 1998.

\bibitem{briegel1998quantum}
H-J Briegel, Wolfgang D{\"u}r, Juan~I Cirac, and Peter Zoller.
\newblock {Quantum repeaters: the role of imperfect local operations in quantum
  communication}.
\newblock {\em Physical Review Letters}, 81(26):5932, 1998.

\bibitem{etxezarreta2021time}
Josu Etxezarreta~Martinez, Patricio Fuentes, Pedro Crespo, and Javier
  Garcia-Frias.
\newblock Time-varying quantum channel models for superconducting qubits.
\newblock {\em npj Quantum Information}, 7(1):1--10, 2021.

\bibitem{kubica2022erasure}
Aleksander Kubica, Arbel Haim, Yotam Vaknin, Fernando Brand{\~a}o, and Alex
  Retzker.
\newblock Erasure qubits: Overcoming the $ t\_1 $ limit in superconducting
  circuits.
\newblock {\em arXiv preprint arXiv:2208.05461}, 2022.

\bibitem{panayi2014memory}
Christiana Panayi, Mohsen Razavi, Xiongfeng Ma, and Norbert L{\"u}tkenhaus.
\newblock Memory-assisted measurement-device-independent quantum key
  distribution.
\newblock {\em New Journal of Physics}, 16(4):043005, 2014.

\bibitem{piparo2017memory}
Nicol{\'o}~Lo Piparo, Mohsen Razavi, and William~J Munro.
\newblock Memory-assisted quantum key distribution with a single
  nitrogen-vacancy center.
\newblock {\em Physical Review A}, 96(5):052313, 2017.

\bibitem{hetet2008quantum}
Gabriel H{\'e}tet.
\newblock {\em Quantum memories for continuous variable states of light in
  atomic ensembles}.
\newblock Australian National University, 2008.

\bibitem{otten2021impacts}
Matthew Otten, Keshav Kapoor, A~Bar{\i}{\c{s}} {\"O}zg{\"u}ler, Eric~T Holland,
  James~B Kowalkowski, Yuri Alexeev, and Adam~L Lyon.
\newblock Impacts of noise and structure on quantum information encoded in a
  quantum memory.
\newblock {\em Physical Review A}, 104(1):012605, 2021.

\bibitem{lim2021}
Charles Ci-Wen Lim, Feihu Xu, Jian-Wei Pan, and Artur Ekert.
\newblock {Security Analysis of Quantum Key Distribution with Small Block
  Length and Its Application to Quantum Space Communications}.
\newblock {\em Physical Review Letters}, 126:100501, Mar 2021.

\end{thebibliography}

\appendix
\section{Detecting amplitude-damping errors}\label{appendix:ancilla-based PS}
We show that the amplitude damping error detection circuit shown in Fig.~\ref{fig:detection} in Subsection~\ref{subsec:dual-rail} works for any arbitrary qubit.
Let the qubit $q_0$ be in an arbitrary state as shown below
\begin{eqnarray}
\rho_{q_0q_1q_2}^{\text{initial}}&=&\vert\alpha\vert^2\ket{000}_{q_0q_1q_2}\bra{000}+\vert\beta\vert^2\ket{100}_{q_0q_1q_2}\bra{100}+\alpha\beta^*\ket{000}_{q_0q_1q_2}\bra{100}\nonumber\\
&&\;\;\;+\alpha^*\beta\ket{100}_{q_0q_1q_2}\bra{000}.
\end{eqnarray}
The dual-rail encoder transforms the physical qubit register to logical qubit register as shown, 
\begin{eqnarray}
\rho_{q_0q_1q_2}^{\text{after encoder}}&=&\vert\alpha\vert^2\ket{010}_{q_0q_1q_2}\bra{010}+\vert\beta\vert^2\ket{100}_{q_0q_1q_2}\bra{100}+\alpha\beta^*\ket{010}_{q_0q_1q_2}\bra{100}\nonumber\\
&&\;\;\;+\alpha^*\beta\ket{100}_{q_0q_1q_2}\bra{010}.
\end{eqnarray}
AD noise corrupts the qubit register,
\begin{eqnarray}
\rho_{q_0q_1q_2}^{\text{after AD}}&=&(1-\gamma)\big(\vert\alpha\vert^2\ket{010}_{q_0q_1q_2}\bra{010}+\vert\beta\vert^2\ket{100}_{q_0q_1q_2}\bra{100}+\alpha\beta^*\ket{010}_{q_0q_1q_2}\bra{100}\nonumber\\
\;\;\;&&+\alpha^*\beta\ket{100}_{q_0q_1q_2}\bra{010}\big)+\gamma\left(\vert\alpha\vert^2+\vert\beta\vert^2\right)\ket{000}_{q_0q_1q_2}\bra{000}.
\end{eqnarray}
Erroneous states are discarded by the post-selection subcircuit with probability $\gamma$ shown below,
\begin{eqnarray}
\rho_{q_0q_1q_2}^{\text{after PS}}&=&(1-\gamma)\Big(\vert\alpha\vert^2\ket{011}_{q_0q_1q_2}\bra{011}+\vert\beta\vert^2\ket{101}_{q_0q_1q_2}\bra{101}+\alpha\beta^*\ket{011}_{q_0q_1q_2}\bra{101}\nonumber\\
&&\;\;\;+\alpha^*\beta\ket{101}_{q_0q_1q_2}\bra{011}\Big).
\end{eqnarray}
Hence, we get the initial state with probability $1-\gamma$ as shown,
\begin{eqnarray}
\rho_{q_0q_1q_2}^{\text{after decoder}}&=&\vert\alpha\vert^2\ket{000}_{q_0q_1q_2}\bra{000}+\vert\beta\vert^2\ket{100}_{q_0q_1q_2}\bra{100}+\alpha\beta^*\ket{000}_{q_0q_1q_2}\bra{100}\nonumber\\
&&\;\;\;+\alpha^*\beta\ket{100}_{q_0q_1q_2}\bra{000}\nonumber \\
&=&(1-\gamma)\rho_{q_0q_1q_2}^{\text{initial}}.
\end{eqnarray}
\section{Effect of AD noise and measurement errors on secure key rates of BB84 and B92 protocol}\label{appendix:AD on BB84 states}
Subsection~\ref{subsec: AD on BB84 key rate analytical} describes the effects of AD noise on the BB84 protocol. Here, we show in detail the effects of AD noise on the four BB84 states. Furthermore, we quantify the effects of qubit read-out errors on the secure key rate of BB84 and B92 protocol.

AD noise has no effect on $\ket{0}$, as shown below
\begin{eqnarray}
\Lambda^{\text{AD}}(\ket{0}\bra{0})&=&\sum_{i=0}^{i=1}A_i^{\text{AD}} \ket{0}\bra{0} A_i^{\dagger\text{AD}} \nonumber \\
&=& \ket{0}\bra{0}.
\label{eq: AD 0}
\end{eqnarray}
AD noise doesnot affect the remaining three BB84 states with probability $1-\gamma$ and maps them to $\ket{0}$ with probability $\gamma$. This is evident from the calculations shown below.
\begin{eqnarray}
\Lambda^{\text{AD}}(\ket{1}\bra{1})&=&\sum_{i=0}^{i=1}A_i^{\text{AD}} \ket{1}\bra{1} A_i^{\dagger\text{AD}} \nonumber \\
&=& (1-\gamma)\ket{1}\bra{1}+\gamma\ket{0}\bra{0}.
\label{eq: AD 1}
\end{eqnarray}

\begin{eqnarray}
\Lambda^{\text{AD}}(\ket{+}\bra{+})&=& \frac{1}{2}\sum_{i=0}^{i=1}\big(A_i^{\text{AD}} \ket{0}\bra{0} A_i^{\dagger\text{AD}}+A_i^{\text{AD}} \ket{0}\bra{1} A_i^{\dagger\text{AD}} +A_i^{\text{AD}} \ket{1}\bra{0}A_i^{\dagger\text{AD}}\nonumber \\ &&\;\;\;\;+A_i^{\text{AD}} \ket{1}\bra{1} A_i^{\dagger\text{AD}}\big)\nonumber \\
&=& \frac{1}{2}\big((1+\gamma)\ket{0}\bra{0} + \sqrt{(1-\gamma)}\ket{0}\bra{1} +\sqrt{(1-\gamma)}\ket{1}\bra{0}\nonumber\\ &&\;\;\;\;+(1-\gamma)\ket{1}\bra{1}\big).
\label{eq: AD +}
\end{eqnarray}

\begin{eqnarray}
\Lambda^{\text{AD}}(\ket{-}\bra{-})&=& \frac{1}{2}\sum_{i=0}^{i=1}\big(A_i^{\text{AD}} \ket{0}\bra{0} A_i^{\dagger\text{AD}}-A_i^{\text{AD}} \ket{0}\bra{1} A_i^{\dagger\text{AD}} -A_i^{\text{AD}} \ket{1}\bra{0}A_i^{\dagger\text{AD}}\nonumber\\
&&\;\;\;+A_i^{\text{AD}} \ket{1}\bra{1} A_i^{\dagger\text{AD}}\big)\nonumber \\
&=& \frac{1}{2}\big((1+\gamma)\ket{0}\bra{0} - \sqrt{(1-\gamma)}\ket{0}\bra{1} -\sqrt{(1-\gamma)}\ket{1}\bra{0}\nonumber\\ \;\;\;&&+(1-\gamma)\ket{1}\bra{1}\big).
\label{eq: AD -}
\end{eqnarray}
Next, we quantify the effects of qubit read-out errors on the security of BB84 protocol. Bob's noisy projective measurement in the $\{\ket{0},\ket{1}\}$ basis is modelled as a generalized measurement (POVM) with positive operators
\begin{eqnarray}
&&P_0=(1-\delta)\ket{0}\bra{0}+\delta\ket{1}\bra{1},\nonumber \\
&&P_1=(1-\delta)\ket{1}\bra{1}+\delta\ket{0}\bra{0},
\label{eq:POVM_noisy_Bob}
\end{eqnarray}
where $0 \leq \delta \leq 1$ quantifies the qubit read-out error. Similarly, we use $P_{\pm}$ to denote Bob's measurement operators for the $X$ basis. The noisy measurement of Bob results in the bit and phase error rates as shown below
\begin{eqnarray}
     &&e_b^\text{{AD}}= \frac{\gamma}{2}-\delta(\gamma -1),\nonumber\\
    &&e_p^{\text{AD}}=\frac{1}{2}\left(1+\sqrt{1-\gamma}(2\delta-1)\right).
 \end{eqnarray}
 Hence, key rate of the BB84 protocol under AD noise and measurement errors is given by,
 \begin{eqnarray}
 R_\text{BB84}^{\text{AD}}&=&1-h\left(\frac{\gamma}{2}-\delta(\gamma -1)\right)-h\left(\frac{1}{2}\left(1+\sqrt{1-\gamma}(2\delta-1)\right)\right).
 \end{eqnarray}
Finally, we show the effects of Bob's imperfect measurement on the secure key rates of B92 protocol. The qubit-readout errors leads to the bit and phase error rates as shown below
\begin{eqnarray}
     &&e_b= \delta,\nonumber\\
         && e_p=\frac{1}{2}\left(1+\sqrt{1-\gamma}(2\delta-1)\right).
 \end{eqnarray}
 Hence, the key rate of B92 under AD noise and qubit-readout error is
 \begin{eqnarray}
 R_\text{B92}&=&1-h(\delta)-h\left(\frac{1}{2}\left(1+\sqrt{1-\gamma}(2\delta-1)\right)\right).
 \end{eqnarray}

\section{Effect of GAD channel on the BB84 states}\label{appendix:GAD on BB84 states}
Subsection~\ref{subsec:GAD on BB84} describes the effects of AD noise on the BB84 protocol. Here, we show in detail the effects of GAD noise on the four BB84 states.
GAD noise maps $\ket{0}$ to $\ket{1}$ with a probability as shown below.
\begin{eqnarray}
\Lambda^{\text{GAD}}(\ket{0}\bra{0})&=&\sum_{i=0}^{i=3}A_i^{\text{GAD}} \ket{0}\bra{0} A_i^{\dagger \text{GAD}}\nonumber \\
&=& [1-(1-p)\gamma]\ket{0}\bra{0}+(1-p)\gamma\ket{1}\bra{1}.
\label{eq: GAD 0}
\end{eqnarray}
Similarly, GAD noise takes $\ket{1}$ to $\ket{0}$ state with a probability as shown below.
\begin{eqnarray}
\Lambda^{\text{GAD}}(\ket{1}\bra{1})&=&\sum_{i=0}^{i=3}A_i^{\text{GAD}} \ket{1}\bra{1} A_i^{\dagger \text{GAD}} \nonumber \\
&=& p\gamma\ket{0}\bra{0}+(1-p\gamma)\ket{1}\bra{1}.
\label{eq: GAD 1}
\end{eqnarray}
Below calculations show how GAD effects $\ket{+}$ and $\ket{- states}$.
\begin{eqnarray}
\Lambda^{\text{GAD}}(\ket{+}\bra{+})&=& \frac{1}{2}\sum_{i=0}^{i=3}\big(A_i^{\text{GAD}} \ket{0}\bra{0} A_i^{\dagger \text{GAD}}+A_i^{\text{GAD}} \ket{0}\bra{1} A_i^{\dagger \text{GAD}}\nonumber \\
&&\;\;\; +A_i^{\text{GAD}} \ket{1}\bra{0} A_i^{\dagger \text{GAD}}+A_i^{\text{GAD}} \ket{1}\bra{1} A_i^{\dagger \text{GAD}}\nonumber \\
&=& (1-\gamma +2p\gamma)\ket{0}\bra{0}+\sqrt{1-\gamma}\ket{0}\bra{1} +\sqrt{1-\gamma}\nonumber\\
&&\;\;\;\times \ket{1}\bra{0}+(1+\gamma-2p\gamma)\ket{0}\bra{0}.
\label{eq: GAD +}
\end{eqnarray}

\begin{eqnarray}
\Lambda^{\text{GAD}}(\ket{-}\bra{-})&=& \frac{1}{2}\sum_{i=0}^{i=3}\big(A_i^{\text{GAD}} \ket{0}\bra{0} A_i^{\dagger \text{GAD}}-A_i^{\text{GAD}} \ket{0}\bra{1} A_i^{\dagger \text{GAD}}\nonumber \\
&&\;\;\; -A_i^{\text{GAD}} \ket{1}\bra{0} A_i^{\dagger \text{GAD}}+A_i^{\text{GAD}} \ket{1}\bra{1} A_i^{\dagger \text{GAD}}\nonumber \\
&=& (1-\gamma +2p\gamma)\ket{0}\bra{0}-\sqrt{1-\gamma}\ket{0}\bra{1} -\sqrt{1-\gamma}\nonumber\\
&&\;\;\;\times \ket{1}\bra{0}+(1+\gamma-2p\gamma)\ket{0}\bra{0}.
\label{eq: GAD -}
\end{eqnarray}

\section{Performance of dual-rail BB84 with post-selection technique against Generalized Amplitude Damping noise}\label{sec:GAD_dual}
In Section~\ref{subsec:optimal PS}, we have described that the error detection circuit of Fig.~\ref{fig:detection} can correct a fraction of GAD errors too. Here, we estimate the fraction of GAD errors detectable by our error detection circuit. We define the Kraus operators of the GAD channel for the dual-rail qubits as
\begin{equation}
    M^{\text{GAD}}_i=A^{\text{GAD}}_j\otimes A^{\text{GAD}}_k,\label{eq:GAD_dual}
\end{equation}
where $j$ and $k$ $\in \{0,1,2,3\}$ and i $\in\{0,1,\dots,15\}$.  
We show the action of Kraus operators of the GAD channel on $\ket{01}$ as
\begin{eqnarray}
\Lambda^{\text{GAD}}(\ket{01}\bra{01})&=&\sum_{i=0}^{15}M^{\text{GAD}}_i\ket{01}\bra{01}M^{\dagger\text{GAD}}_i\nonumber\\
&=&
p\gamma\left(1-\gamma-p\gamma\right)\ket{00}\bra{00}+\left(1-\gamma+p\gamma^2-\gamma^2p^2\right)\nonumber\\
&&\times\ket{01}\bra{01}+p\gamma^2\left(1-p\right)\ket{10}\bra{10}\nonumber\\
&&+\gamma\left(1-p-p\gamma+p^2\gamma\right)\ket{11}\bra{11}.
\label{eq:GAD_01}\end{eqnarray}
Along the similar lines we apply the Kraus operators for the dual-rail qubits (see Eq.~\eqref{eq:GAD_dual}) to the remaining BB84 encoded states, $\ket{1}_L$, $\ket{+}_L$ and $\ket{-}_L$, and obtain bit and phase error rates as 
\begin{equation}
    e_b=e_p=\frac{p\gamma^2(1-p)}{1-\gamma+2p\gamma^2-2p^2\gamma^2}\label{eq:GAD_error_dual}
\end{equation}
We also obtain the probability of error detection, which leads to a sifted key rate of
\begin{equation}
    R_{\text{sift}}=1-\gamma+2p\gamma^2-2p^2\gamma^2.
\end{equation}

\section{Dual-rail BB84 with optimal post-selection technique}\label{Appendix: arbitrary state optimal PS}
In Subsection~\ref{subsec:optimal PS}, we present a dual-rail encoded BB84 circuit with minimal number of \textsc{cnot}s. Here, we show that such an optimal post-selection circuit can detect AD errors for any arbitrary qubit state. Suppose, the qubit register is in an arbitrary state as shown below
\begin{equation}
\rho_{q_0q_1}^{\text{initial}}=\vert\alpha\vert^2\ket{00}_{q_0q_1}\bra{00}+\vert\beta\vert^2\ket{10}_{q_0q_1}\bra{10}+\alpha\beta^*\ket{00}_{q_0q_1}\bra{10}+\alpha^*\beta\ket{10}_{q_0q_1}\bra{00}
\end{equation}
The dual-rail encoder works as
\begin{eqnarray}
\rho_{q_0q_1}^{\text{after encoder}}&=&\vert\alpha\vert^2\ket{01}_{q_0q_1}\bra{01}+\vert\beta\vert^2\ket{10}_{q_0q_1}\bra{10}+\alpha\beta^*\ket{01}_{q_0q_1}\bra{10}\nonumber\\
&&\;\;\;+\alpha^*\beta\ket{10}_{q_0q_1}\bra{01}.
\end{eqnarray}
AD noise affects this arbitrary dual-rail encoded state as
\begin{eqnarray}
\rho_{q_0q_1}^{\text{after AD}}&=&(1-\gamma)\big(\vert\alpha\vert^2\ket{01}_{q_0q_1}\bra{01}+\vert\beta\vert^2\ket{10}_{q_0q_1}\bra{10}+\alpha\beta^*\ket{01}_{q_0q_1}\bra{10}\nonumber\\
&&+\alpha^*\beta\ket{10}_{q_0q_1}\bra{01}\big)+\gamma\left(\vert\alpha\vert^2+\vert\beta\vert^2\right)\ket{00}_{q_0q_1}\bra{00}.
\end{eqnarray}
The \textsc{cnot} gate (post-selection subcircuit) works as
\begin{eqnarray}
\rho_{q_0q_1}^{\text{after PS}}&=&(1-\gamma)\big(\vert\alpha\vert^2\ket{01}_{q_0q_1}\bra{01}+\vert\beta\vert^2\ket{10}_{q_0q_1}\bra{10}+\alpha\beta^*\ket{01}_{q_0q_1}\bra{10}\nonumber\\
\;\;\;&&+\alpha^*\beta\ket{10}_{q_0q_1}\bra{01}\big)\nonumber\\
&=&(1-\gamma)\rho_{q_0q_1}^{\text{initial}}.
\end{eqnarray}

\end{document}